\documentclass[sigconf]{acmart}
\pdfoutput=1

\usepackage{blindtext, graphicx, xcolor, listings, booktabs, multirow, colortbl, bm, subcaption}
\usepackage{float, placeins}

\usepackage{multicol}

\makeatletter
\newcommand\notsotiny{\@setfontsize\notsotiny{6.0}{7.0}}
\newcommand\notsosmall{\@setfontsize\notsosmall{6.5}{7.5}}
\makeatother

\usepackage[normalem]{ulem}

\usepackage{enumitem}

\usepackage[frozencache]{minted}
\usepackage{tcolorbox}
\tcbuselibrary{minted,skins}
\newtcblisting{stevemint}{
  listing engine=minted,
  colback=gray!10,
  colframe=gray!50,
  listing only,
  minted style=colorful,
  minted language=nasm,
  minted options={linenos=true,texcl=true,fontsize=\footnotesize},
  left=1mm,
}
\definecolor{stevemintrgb}{rgb}{0.85,0.85,0.85}

\usepackage{esvect}

\usepackage{fixltx2e}

\usepackage{makecell}

\usepackage{arydshln}
\usepackage{tcolorbox}

\def\fauxschelper#1 #2\relax{%
\fauxschelphelp#1\relax\relax%
\if\relax#2\relax\else\ \fauxschelper#2\relax\fi%
}
\def\Hscale{.85}\def\Vscale{.74}\def\Cscale{1.12}
\def\fauxschelphelp#1#2\relax{%
\ifnum`#1>``\ifnum`#1<`\{\scalebox{\Hscale}[\Vscale]{\uppercase{#1}}\else%
\scalebox{\Cscale}[1]{#1}\fi\else\scalebox{\Cscale}[1]{#1}\fi%
\ifx\relax#2\relax\else\fauxschelphelp#2\relax\fi}

\usepackage{array}
\usepackage{tabularx}
\newcolumntype{P}[1]{>{\centering\arraybackslash}p{#1}}
\newcolumntype{M}[1]{>{\centering\arraybackslash}m{#1}}

\usepackage{caption}
\definecolor{light-gray}{gray}{0.92}
\definecolor{darkorange}{RGB}{255,100,0}

\usepackage{mdframed}
{\bfseries}
\newmdtheoremenv[%
linecolor=gray,leftmargin=10,
rightmargin=10,
backgroundcolor=gray!16,%
innertopmargin=4pt,%
]{myframe}{Instrumenter Criterion}

\def\mystrut{\rule{0pt}{1\normalbaselineskip}}

\usepackage{pifont}
\newcommand{\cmark}{\ding{52}}%
\newcommand{\xmark}{\ding{55}}%
\newcommand{\xmarktab}{\ding{55}}%

\usetikzlibrary{arrows.meta, shapes.geometric, calc, shadows, chains}
\colorlet{col1in}{gray!10}
\colorlet{col1out}{gray!10}
\colorlet{col2in}{teal!10}
\colorlet{col2out}{teal!10}
\colorlet{col3in}{orange!20}
\colorlet{col3out}{orange!20}
\pgfkeys{
    orect/.append style  = {rect, font = \ttfamily, text width = 325pt, text centered,minimum height = 10pt},
}

\usepackage{footnote}
\makesavenoteenv{tabular}
\makesavenoteenv{table}

\usepackage{pict2e}
\makeatletter
\newcommand{\crossout}[1]{%
  \begingroup
  \settowidth{\dimen@}{#1}%
  \setlength{\unitlength}{0.05\dimen@}%
  \settoheight{\dimen@}{#1}%
  \count@=\dimen@
  \divide\count@ by \unitlength
  \begin{picture}(0,0)
  \put(0,\count@){\line(20,-\count@){20}}
  \end{picture}%
  #1%
  \endgroup
}
\makeatother

\usepackage{amsmath}
\makeatletter
\newsavebox\myboxA
\newsavebox\myboxB
\newlength\mylenA
\newcommand*\xoverline[2][0.75]{%
    \sbox{\myboxA}{$\m@th#2$}%
    \setbox\myboxB\null
    \ht\myboxB=\ht\myboxA%
    \dp\myboxB=\dp\myboxA%
    \wd\myboxB=#1\wd\myboxA
    \sbox\myboxB{$\m@th\overline{\copy\myboxB}$}
    \setlength\mylenA{\the\wd\myboxA}
    \addtolength\mylenA{-\the\wd\myboxB}%
    \ifdim\wd\myboxB<\wd\myboxA%
       \rlap{\hskip 0.5\mylenA\usebox\myboxB}{\usebox\myboxA}%
    \else
        \hskip -0.5\mylenA\rlap{\usebox\myboxA}{\hskip 0.5\mylenA\usebox\myboxB}%
    \fi}
\makeatother

\lstdefinestyle{steve}{
  captionpos=b,
  language=C,
  emptylines=0,
  breaklines=true,
  basicstyle=\footnotesize,
  moredelim=**[is][\color{red}]{@}{@},
  moredelim=**[is][\color{teal}]{@@}{@@},
}

\NewDocumentCommand{\stevebox}{O{}m+m}{%
  \begin{tcolorbox}[%
    sharp corners,
    colframe=gray!50, 
    boxsep = 0pt,
    left = 7pt,
    right = 7pt,
    top = 5pt,
    bottom = 5pt,
    colback=orange!7, 
    box align=center,
    valign=center,
    title = {},#1]
    #3%
    \end{tcolorbox}%
}

\usepackage[ruled,vlined,linesnumbered,noresetcount]{algorithm2e}
\usepackage{algpseudocode}

\newcommand{\eat}[1]{}
\definecolor{purple}{rgb}{1,0,1}

\usepackage{hyperref}
\usepackage{cleveref}
\definecolor{winered}{rgb}{0.5,0,0}
\hypersetup{colorlinks = true, 
    urlcolor = black, 
    linkcolor = winered, 
    citecolor = winered 
}

\usepackage{dblfloatfix}

\newcommand{\platname}{\textsc{HeXcite}\xspace}
\newcommand{\platnameit}{\emph{\textsc{HeXcite}}\xspace}

\definecolor{vt}{HTML}{892242}

\SetCommentSty{mycommfont}

\copyrightyear{2021}
\acmYear{2021}
\setcopyright{acmcopyright}
\acmConference[CCS '21] {Proceedings of the 2021 ACM SIGSAC Conference on Computer and Communications Security}{November 15--19, 2021}{Virtual Event, Republic of Korea.}
\acmBooktitle{Proceedings of the 2021 ACM SIGSAC Conference on Computer and Communications Security (CCS '21), November 15--19, 2021, Virtual Event, Republic of Korea}
\acmPrice{15.00}
\acmISBN{978-1-4503-8454-4/21/11}
\acmDOI{10.1145/3460120.3484787}
\begin{document}
\fancyhead{}

\title{Same Coverage, Less Bloat: Accelerating Binary-only Fuzzing\\with Coverage-preserving Coverage-guided Tracing}

\author{Stefan Nagy}
\affiliation{%
  \institution{Virginia Tech}
  \city{Blacksburg}
  \country{Virginia}}
\email{snagy2@vt.edu}

\author{Anh Nguyen-Tuong}
\affiliation{%
  \institution{University of Virginia}
  \city{Charlottesville}
  \country{Virginia}}
\email{nguyen@virginia.edu}

\author{Jason D. Hiser}
\affiliation{%
  \institution{University of Virginia}
  \city{Charlottesville}
  \country{Virginia}}
\email{hiser@virginia.edu}

\author{Jack W. Davidson}
\affiliation{%
  \institution{University of Virginia}
  \city{Charlottesville}
  \country{Virginia}}
\email{jwd@virginia.edu}

\author{Matthew Hicks}
\affiliation{%
  \institution{Virginia Tech}
  \city{Blacksburg}
  \country{Virginia}}
\email{mdhicks2@vt.edu}


\begin{abstract}
Coverage-guided fuzzing's aggressive, high-volume testing has helped reveal tens of thousands of software security~flaws.
While executing billions of test cases mandates fast code coverage tracing, the nature of \emph{binary-only} targets leads to reduced tracing performance.
A recent advancement in binary fuzzing performance is \emph{Coverage-guided Tracing} (CGT), which brings orders-of-magnitude gains in throughput by restricting the expense of coverage tracing to only when new coverage is guaranteed.
Unfortunately, CGT suits only a basic block coverage granularity---yet most fuzzers require finer-grain coverage metrics: \emph{edge coverage} and \emph{hit counts}. 
It is this limitation which prohibits nearly all of today's state-of-the-art fuzzers from attaining the performance benefits of CGT.

This paper tackles the challenges of adapting CGT to fuzzing's most ubiquitous coverage metrics.
We introduce and implement a suite of enhancements that expand CGT's introspection to fuzzing's most common code coverage metrics, while maintaining its orders-of-magnitude speedup over conventional always-on coverage tracing.
We evaluate their trade-offs with respect to fuzzing performance and effectiveness across \textbf{12} diverse real-world binaries (\textbf{8} open- and \textbf{4} closed-source).
On average, our \emph{coverage-preserving} CGT attains \textbf{near-identical} speed to the present \emph{block-coverage-only} CGT, UnTracer; and outperforms leading binary- \emph{and} source-level coverage tracers QEMU, Dyninst, RetroWrite, and AFL-Clang by \textbf{2--24$\times$}, finding more bugs in less time.

\end{abstract}


\maketitle

{\fontsize{8pt}{8pt} \selectfont
\textbf{ACM Reference Format:}\
Stefan Nagy, Anh Nguyen-Tuong, Jason D. Hiser, Jack W. Davidson, and Matthew Hicks. 2021. Same Coverage, Less Bloat: Accelerating Binary-only Fuzzing with Coverage-preserving Coverage-guided Tracing. In {\it Proceedings of the 2021 ACM SIGSAC Conference on Computer and Communications Security (CCS’21), November 15--19, 2021, Virtual Event, Republic of Korea.} ACM, New York, NY, USA, 15 pages. \newline https://doi.org/10.1145/3460120.3484787 }

\section{Introduction}
\emph{Coverage-guided fuzzing} has become one of the most popular and successful techniques for software security auditing. 
Its aggressive, high-volume testing strategy has revealed countless security vulnerabilities in software, and helped proactively secure many of the world's most popular codebases~\cite{bounimova_billions_2012}. 
Today, software projects of all sizes rely on fuzzing to root out bugs and vulnerabilities both throughout and beyond the software development cycle.

Fuzzing consists of three main steps: (1) \emph{test case generation}, (2) \emph{code coverage tracing}, and (3) \emph{test case triage}.
Many works improve fuzzing at the generation level by incorporating input grammars~\cite{godefroid_grammar-based_2008}, path prioritization~\cite{lemieux_fairfuzz_2018}, better mutators~\cite{lv_mopt_2019}, or constraint solving~\cite{aschermann_redqueen_2018}; while others focus on refining triage with sanitizers~\cite{dinesh_retrowrite_2020} or other heuristics.
However, given fuzzing's core goal of producing---and eventually executing---a large volume of test cases, maintaining high-performance test case execution is critical to effective fuzzing.
Recent work shows both ``dumb'' and ``smart'' fuzzers spend the majority of their time executing test cases and collecting their coverage traces~\cite{nagy_full-speed_2019}. 
However, in binary-only fuzzing contexts, the semantically-poor and opaque nature of a binary prevents the tight integration of coverage-tracing routines that is possible in source-available contexts.
This inflates the tracing overhead by up to two orders of magnitude compared to compiler-based instrumentation of source code\eat{~\cite{nagy_full-speed_2019}}.
Even in an ideal world where black-box instrumenters approach compiler-level performance, recent work shows that coverage tracing increases test case execution time by roughly 30\%~\cite{dinesh_retrowrite_2020}.
To address this performance gap and the time wasted by needless coverage tracing, many binary-only fuzzing efforts~\cite{jung_winnie_2021, gros_fuzzing_2020, toepfer_bsod_2021, fioraldi_afl_2020} are eschewing conventional always-on coverage tracing for an on-demand tracing strategy known as \emph{Coverage-guided Tracing} (CGT)~\cite{nagy_full-speed_2019}.
CGT restricts the expense of tracing to only when new coverage is guaranteed (roughly 1 test case in every 10,000), thereby increasing fuzzing throughput by 500--600\% over the leading binary-only tracers.

While some practitioners are leveraging the idea of CGT~\cite{gros_fuzzing_2020, jung_winnie_2021, fioraldi_afl_2020}, we are not aware of any fuzzer that has adopted it as a replacement for always-on tracing.
Our survey of 27 fuzzers reveals why such performance benefits sit unrealized: CGT only supports \emph{basic block} coverage (instruction sequences ending in control-flow transfer), but most fuzzers rely on finer-grained coverage metrics.
Specifically, our study shows \textbf{25/27} adopt \emph{edges} (transitions between blocks), and \textbf{26/27} further track block or edge \emph{hit counts} (execution frequencies).
This lack of support for the most common coverage metrics inhibits CGT's adoption in nearly all fuzzers.
While CGT's near-0\% tracing overhead is ideal for fuzzing's high-throughput needs, its coverage deficiencies force today's state-of-the-art fuzzers to instead rely on orders-of-magnitude slower, always-on tracing---leaving their full performance potential unrealized.

This paper tackles the challenge of extending CGT to fuzzing's most ubiquitous coverage metrics---edges and hit counts---making high-performance tracing available for all existing (and future) fuzzers. 
At the core of our efforts are binary-level and fuzzing enhancements that broaden CGT's coverage while maintaining its orders-of-magnitude speedup:
for edge coverage, we introduce a zero-overhead strategy called \emph{jump mistargeting} that addresses the most common (statically and dynamically) form of critical edges while keeping control flow intact.
To maintain completeness of edge coverage, we back jump mistargeting with a low-overhead binary-only implementation of a control-flow transformation that eliminates critical edges through block insertion called branch splitting (e.g., LLVM's SanitizerCoverage~\cite{the_clang_team_sanitizercoverage_2019}).
To extend CGT to hit count coverage, we exploit the observation that execution frequency changes are highly localized to loops, devising a \emph{bucketed unrolling} strategy to encode them with a minimally-invasive hit count tracking mechanism congruent with current fuzzers~\cite{zalewski_american_2017, fioraldi_afl_2020}.

We implement our \emph{coverage-preserving} Coverage-guided Tracing, \platnameit, and evaluate it against the current block-coverage-only CGT implementation UnTracer~\cite{nagy_full-speed_2019}; the leading binary-only fuzzing coverage tracers QEMU~\cite{zalewski_american_2017}, Dyninst~\cite{heuse_afl-dyninst_2018}, and RetroWrite~\cite{dinesh_retrowrite_2020}; and the popular source-level coverage tracing via AFL-Clang~\cite{zalewski_american_2017}.
In evaluations across \textbf{12} diverse real-world binaries (\textbf{8} open- and \textbf{4} closed-source), \platname attains a throughput near identical to UnTracer's; \textbf{3--24$\times$} that of conventional \emph{always-on} binary-only tracers QEMU, Dyninst, and RetroWrite; and \textbf{2.8$\times$} that of source-level tracing with AFL-Clang.
\platname's coverage-preserving transformations further enable it to find \textbf{12--749\%} more unique bugs than UnTracer as well as always-on binary- \emph{and} source-level tracers in standard coverage-guided grey-box fuzzing integrations---while finding \textbf{16} known bugs and vulnerabilities in \textbf{32--52\%} less time.

Through the following contributions, \textbf{this paper enables the use of the fastest tracing approach in fuzzing---\emph{Coverage-guided Tracing}---by the majority of today's fuzzers}:

\begin{itemize}
\setlength\itemsep{-0.0em}
    \item We introduce \emph{jump mistargeting}: a control-flow redirection strategy which alters the common-case of edge instructions such that they self-report edge coverage at native speed.
    \item We introduce \emph{bucketed unrolling}: a technique which clones loop conditions at discrete intervals, enabling the self-reporting of loop hit-count coverage at near-native speed. 
    \item We demonstrate that with these techniques, our \emph{coverage-preserving} CGT eclipses block-only CGT---and conventional always-on binary- \emph{and} source-level tracers---in edge coverage, loop coverage, and bug-finding fuzzing effectiveness. 
    \item We show that coverage-preserving CGT's speed is nearly indistinguishable from that of block-only CGT, and---despite being a binary-only technique---is >2$\times$ the speed of even source-level tracing approaches. 
    \item We open-source \platname, our implementation of binary-only coverage-preserving CGT, and our evaluation benchmarks at: \texttt{https://github.com/FoRTE-Research/HeXcite}.
\end{itemize}

\section{Background}
To understand our improvements to Coverage-guided Tracing, it is crucial to understand the core details of coverage-guided fuzzing, its code coverage metrics, and the high-performance tracing strategy known as Coverage-guided Tracing. 

\subsection{Software Fuzzing}
Software fuzzing broadly represents one of today's most popular software quality assurance approaches.
Unlike other forms of software testing that vet functionality (e.g., unit testing, mutational testing), fuzzing's primary focus is security auditing;
test cases are generated and fed to the target program with their effects monitored for signs of security violations.
Many software vulnerabilities have been (and continue to be) uncovered via fuzzing, and its use among developers large and small continues to grow each year~\cite{fioraldi_afl_2020}. 

Fuzzing encompasses a variety of techniques accommodating specific use cases, with the most common distinction being search strategy;
\emph{directed} fuzzers constrain testing to specific code or paths (e.g., newly-patched~\cite{bohme_directed_2017} or likely-vulnerable code~\cite{chen_savior_2020}), while \emph{guided} fuzzers aim to maximize the program's state space along some pre-specified metric (e.g., memory accesses or code coverage).
By far the most common and successful form of fuzzing is \emph{coverage-guided fuzzing}~\cite{zalewski_american_2017} which, as the name implies, aims to maximize test cases' code coverage to uncover hidden program bugs.

\subsection{Coverage-guided Fuzzing}
\label{sec:back:cgf}

Coverage-guided fuzzing's scalability, easy adoption, and time-tested effectiveness have made it widely popular among both developers and security practitioners.
As shown in \autoref{fig:cggbf}, given a target program, a typical coverage-guided fuzzing workflow consists of the following recurring steps:

\vspace{-0.1cm}
\begin{figure}[!h]
    \centering
    \framebox{\includegraphics[width=0.94\linewidth, trim={4ex 3ex 8ex 2ex},clip]{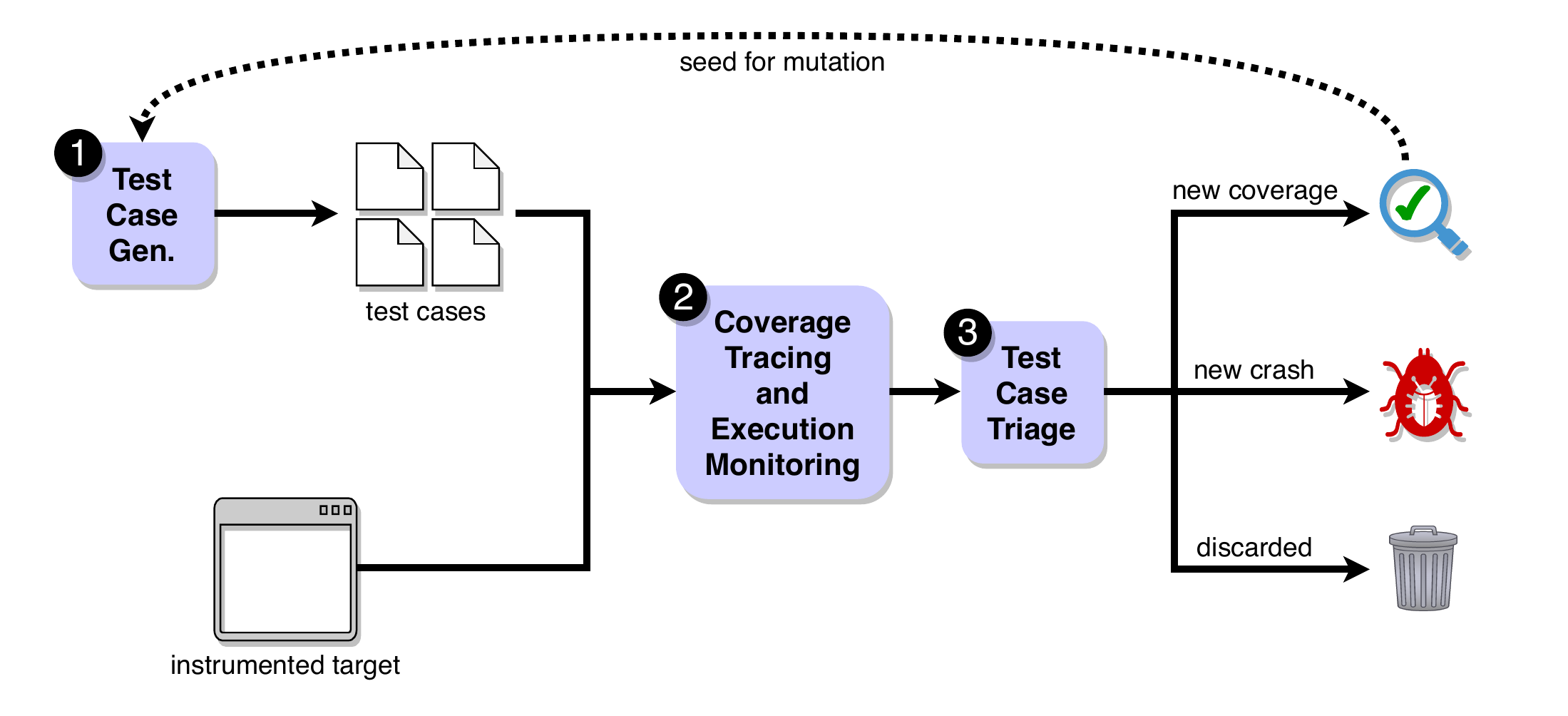}}
    \vspace{-0.35cm}
    \caption{The high-level steps of coverage-guided fuzzing.}
    \label{fig:cggbf}
    \vspace{-0.25cm}
    \hrulefill
\end{figure}

\begin{enumerate}[]
\setlength\itemsep{-0.0em}
    \item \textbf{\emph{Generation}}. Genetic algorithms (typically a mix of random and deterministic byte mutations) create batches of candidate test cases from one or more ancestors. 
    \item \textbf{\emph{Coverage Tracing \& Execution Monitoring}}. Lightweight statically- or dynamically-inserted instrumentation captures each test case's runtime code coverage given some pre-specified coverage metric(s), while monitoring their other execution behavior (e.g., terminating signal).
    \item \textbf{\emph{Triage}}. Candidates are grouped based on observed execution behavior; those increasing coverage are preserved for future mutation, while those triggering crashes are deduplicated in anticipation of manual bug analysis.
\end{enumerate}

Coverage-guided fuzzing's balance of feedback-guided auditing and aggressive, high-volume testing continues to reign supreme over other automated security testing methodologies; 
its effectiveness is evidenced by the deep (and ever-growing) vulnerability trophy cases held by prominent fuzzers such as Google's AFL~\cite{zalewski_american_2017}, honggFuzz~\cite{swiecki_honggfuzz_2018}, and libFuzzer~\cite{serebryany_continuous_2016}; 
and its fundamental principles form the core of today's most state-of-the-art fuzzing efforts.

\subsection{Fuzzing's Code Coverage Metrics}
\label{sec:back:covg}
To maximally vet the target application, coverage-guided fuzzing collects a test case's dynamic code coverage and subsequently mutates only those which attain new coverage.
In our efforts to understand fuzzing's current coverage landscape, we survey 27 of today's state-of-the-art coverage-guided fuzzers (\autoref{tab:survey}) and identify three universal coverage metrics: \emph{basic blocks}, \emph{edges}, and \emph{hit counts}. 
We discuss these coverage metrics in detail below.

\begin{table}[!h]
\scriptsize
    
    \centering
    
    \begin{tabular}{p{1.16cm} p{0.25cm}p{0.25cm} | p{1.22cm} M{0.25cm}M{0.25cm} | p{1.16cm} M{0.25cm}M{0.25cm} }\specialrule{.1em}{0em}{0em}\hline
        \textbf{Name} & \textbf{Cov} & \textbf{Hit} & \textbf{Name} & \textbf{Cov} & \textbf{Hit} & \textbf{Name} & \textbf{Cov} & \textbf{Hit} \\ 
        \hline\hline
        \rowcolor{gray!10} AFL~\cite{zalewski_american_2017}      & \ding{228}& \cmark & EnFuzz~\cite{chen_enfuzz_2019}     & \ding{228}& \cmark & ProFuzzer~\cite{you_profuzzer_2019} & \ding{228}& \cmark \\
        AFL++~\cite{fioraldi_afl_2020}    & \ding{228}& \cmark & FairFuzz~\cite{lemieux_fairfuzz_2018}   & \ding{228}& \cmark & QSYM~\cite{yun_qsym_2018}      & \ding{228}& \cmark \\
        \rowcolor{gray!10} AFLFast~\cite{bohme_coverage-based_2016}  & \ding{228}& \cmark & honggFuzz~\cite{swiecki_honggfuzz_2018}  & \ding{228}& \xmark & REDQUEEN~\cite{aschermann_redqueen_2018}  & \ding{228}& \cmark \\
        AFLSmart~\cite{pham_smart_2019} & \ding{228}& \cmark & GRIMORE~\cite{blazytko_grimoire_2019}    & \ding{228}& \cmark & SAVIOR~\cite{chen_savior_2020}    & \ding{228}& \cmark \\
        \rowcolor{gray!10} Angora~\cite{chen_angora_2018}   & \ding{228}& \cmark & lafIntel~\cite{noauthor_laf-intel_2016}  & \ding{228}& \cmark & SLF~\cite{you_slf_2019}       & \ding{228}& \cmark \\
        CollAFL~\cite{gan_collafl_2018}  & \ding{228}& \cmark & libFuzzer~\cite{serebryany_continuous_2016}  & \ding{228}& \cmark & Steelix~\cite{li_steelix_2017}   & \ding{228}& \cmark \\ 
        \rowcolor{gray!10} DigFuzz~\cite{zhao_send_2019}  & \ding{228}& \cmark & Matryoshka~\cite{chen_matryoshka_2019} & \ding{228}& \cmark & Superion~\cite{wang_superion_2019}  & \ding{228}& \cmark \\ 
        Driller~\cite{stephens_driller_2016}  & \ding{228}& \cmark & MOpt~\cite{lv_mopt_2019}   & \ding{228}& \cmark & TIFF~\cite{jain_tiff_2018}    & \ding{110} & \cmark \\ 
        \rowcolor{gray!10} Eclipser~\cite{choi_grey-box_2019} & \ding{228}& \cmark & NEUZZ~\cite{she_neuzz_2019}      & \ding{228}& \cmark & VUzzer~\cite{rawat_vuzzer_2017}      & \ding{110} & \cmark \\ 
        \hline
    \end{tabular}
    \vspace{0.1cm}
    \caption{A survey of recent coverage-guided fuzzers and their coverage metrics (edges/blocks and hit counts). Key: \ding{228} (edges), \ding{110} (blocks).}
    \label{tab:survey}
\vspace{-0.4cm}
\hrulefill{}
\end{table}

\par\textbf{Basic Block Coverage:}
Basic blocks refer to straight-line (i.e., single entry and exit) instruction sequences beginning and ending in control-flow transfer (i.e., jumps, calls, or returns), and comprise the nodes of a program's control-flow graph.
Tracking basic block coverage necessitates instrumenting each to record their execution in some data structure (e.g., an array~\cite{rawat_finding_2012} or bitmap~\cite{zalewski_american_2017}).
Two modern fuzzers that employ basic block coverage are VUzzer~\cite{rawat_vuzzer_2017} and its successor TIFF~\cite{jain_tiff_2018}.

\par\textbf{Edge Coverage:}
Edges refer to block-to-block transitions, and offer a finer-grained approximation of paths taken.
As \autoref{tab:survey} shows, most fuzzers rely on edge coverage;
AFL~\cite{zalewski_american_2017} and its many derivatives~\cite{fioraldi_afl_2020} record edges as hashes of their start/end block tuples in a bitmap data structure; while LLVM SanitizerCoverage-based~\cite{the_clang_team_sanitizercoverage_2019} fuzzers honggFuzz and libFuzzer track edges from the block level by splitting \emph{critical edges} (edges whose start/end blocks have at least two outgoing/incoming edges, respectively).

\par\textbf{Hit Count Coverage:}
\label{sec:back:covg:hit}
Hit counts refer to block/edge execution frequencies, and are commonly tracked to reveal progress in state exploration (e.g., iterating on a loop).
libFuzzer, AFL, and AFL derivatives approximate hit counts using 8-bit ``buckets'', with each bit representing one of eight ranges (0--1, 2, 3, 4--7, 8--15, 16--31, 32--127, 128+); test cases are seeded if they jump from one bucket to another for any block/edge.
While tracking \emph{exact} hit counts (e.g., VUzzer and TIFF) reveals finer-grained state changes, it risks over-saturating the fuzzer seed pool with needless test cases (e.g., one per new loop iteration), and is hence seldom used. 

\subsection{Coverage-guided Tracing}
Many recent works improve fuzzing with smarter test case generation or triage. 
But despite these advancements, the maximal performance of both standard and state-of-the-art coverage-guided fuzzers is subject to a key constraint: code coverage is traced for \emph{all} test cases, yet \emph{less than 1 in 10,000} actually increase coverage~\cite{nagy_full-speed_2019}.
While this has little impact in use cases where tracing instrumentation is already fast (i.e., open-source software), it is the principal bottleneck for those where tracing is costly---i.e., closed-source software.
For this reason, a number of binary-only fuzzing efforts~\cite{gros_fuzzing_2020, jung_winnie_2021, toepfer_bsod_2021, fioraldi_afl_2020} are instead adopting a lighter-weight strategy called \emph{Coverage-guided Tracing} (CGT), which restricts the expense of tracing to only the $<0.01\%$ of test cases that increase coverage.

\vspace{-0.1cm}
\begin{figure}[h]
    \centering
    \framebox{\includegraphics[width=0.97\linewidth, trim={5.75ex 3.7ex 5.1ex 3.8ex},clip]{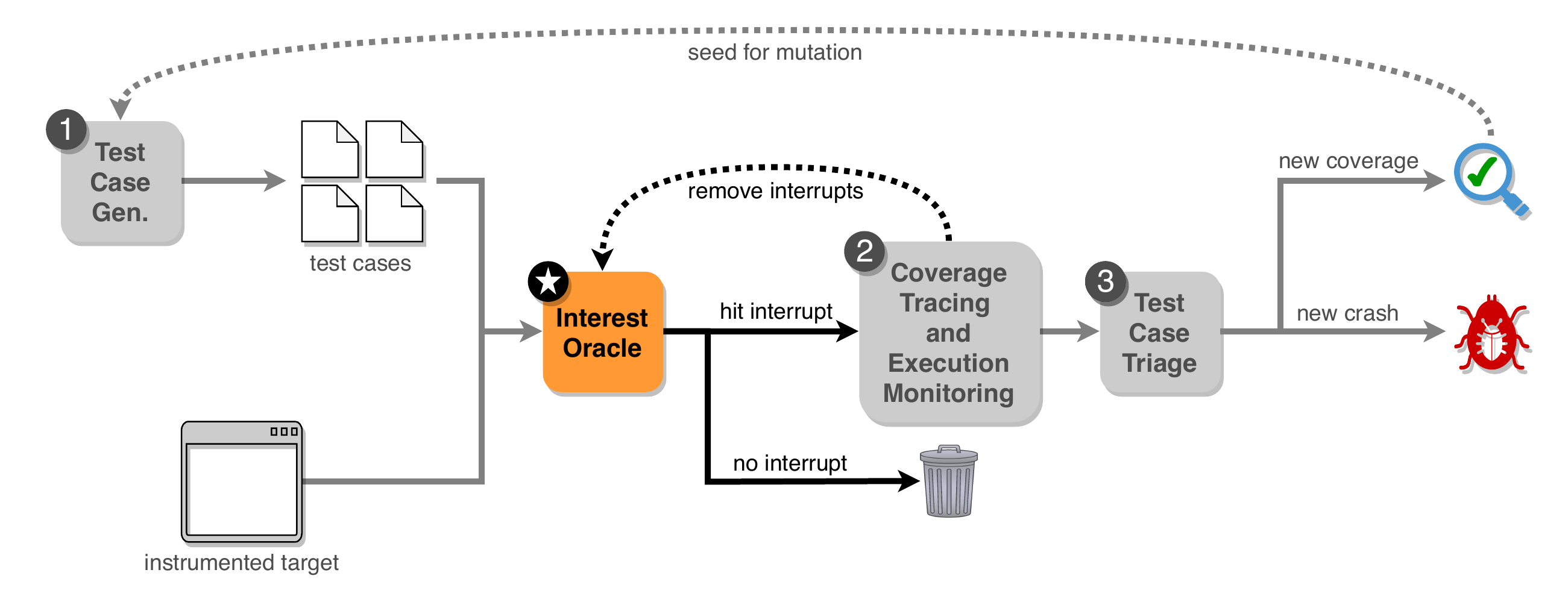}}
    
    \vspace{-0.35cm}
    \caption{Coverage-guided Tracing's core workflow.}
    \label{fig:cgt}
    \vspace{-0.25cm}
    \hrulefill
\end{figure}

Given a target binary, CGT constructs two versions: a \emph{coverage oracle} with an interrupt (e.g., \texttt{0xCC}) inserted at every basic block, and a \emph{tracer} instrumented for conventional fuzzing coverage tracing.
As shown in \autoref{fig:cgt}, CGT runs each test case first on the oracle; if an interrupt is hit, the test case's full coverage is then captured with the tracer, and all visited blocks' have their corresponding oracle interrupts removed; and if no interrupt was hit, the test case is simply discarded following its run on the oracle.
Most test cases ($>99.9\%$~\cite{nagy_full-speed_2019}) revisit already-seen coverage and thus will not trigger interrupts, sparing them of tracing; and because the oracle's mechanism of reporting new coverage is just interrupts (and not instrumentation callbacks) this majority of test cases are run at speed equivalent to the original binary's---giving CGT a near-native runtime overhead of $0.3\%$, and 500--600\% higher test case throughput over the conventional \emph{always-on} tracing used in binary-only fuzzing like AFL-Dyninst~\cite{heuse_afl-dyninst_2018} and AFL-QEMU~\cite{zalewski_american_2017}.

\vspace{0.1cm}
\stevebox{}{
\small
\textbf{The Code Coverage Dilemma:} Though CGT enables orders-of-magnitude higher binary-only fuzzing throughput, it is currently incompatible with \textbf{all} of the state-of-the-art coverage-guided fuzzers we surveyed in \autoref{tab:survey}:
whereas CGT presently supports only a basic block coverage level, \textbf{25/27} fuzzers instead rely on edge coverage, and \textbf{26/27} further track hit counts.
Allowing the broad spectrum of coverage-guided fuzzers to obtain the performance benefits of CGT necessitates an answer to this disparity in code coverage metrics.
}

\section{A Coverage-preserving CGT}
Coverage-guided Tracing (CGT) accelerates binary-only fuzzing by restricting the expense of code coverage tracing to only the few test cases that reach new coverage.
Unfortunately, CGT's lack of support for fuzzing's most common coverage metrics, edges and hit counts, leaves its performance benefits untapped for nearly all of today's state-of-the-art fuzzers.

To address this incompatibility, we observe how CGT achieves lightweight coverage tracking at the control-flow level; and devise two new techniques exploiting this paradigm to facilitate finer-grained coverage---\emph{jump mistargeting} (for edge coverage) and \emph{bucketed unrolling} (for hit counts)---without compromising CGT's minimally-invasive nature.
Below we discuss the inner workings of jump mistargeting and bucketed unrolling, and the underlying insights and observations that motivate them. 

\begin{figure}[H]
    \centering
    \tiny
    \begin{subfigure}[t]{0.236\textwidth}
          \includegraphics[width=\linewidth]{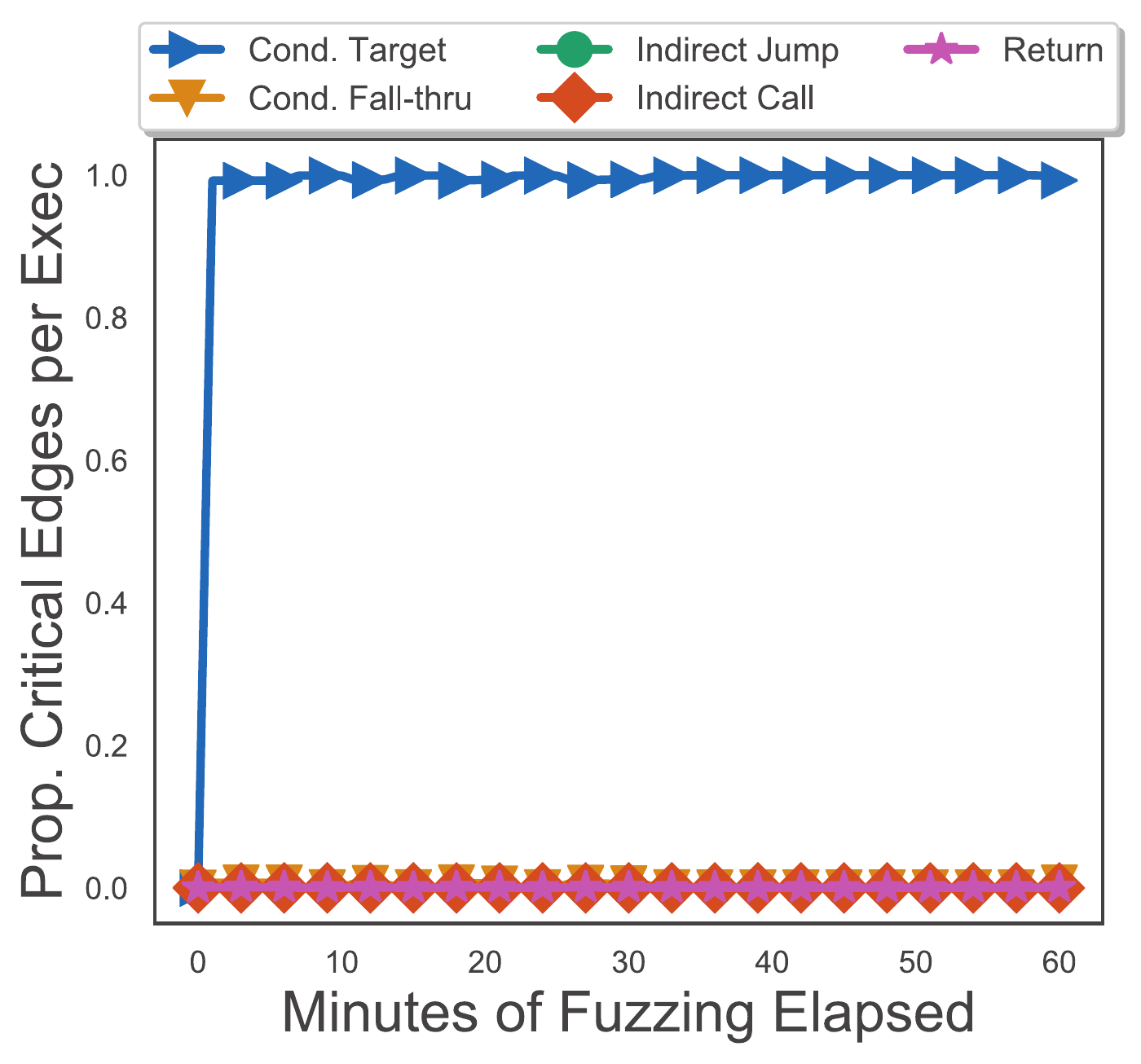}
          \vspace{-0.45cm}
          \caption{ \texttt{bsdtar}}
          \label{fig:critedgesdynamic:A}
     \end{subfigure}
     \begin{subfigure}[t]{0.236\textwidth}
          \includegraphics[width=\linewidth]{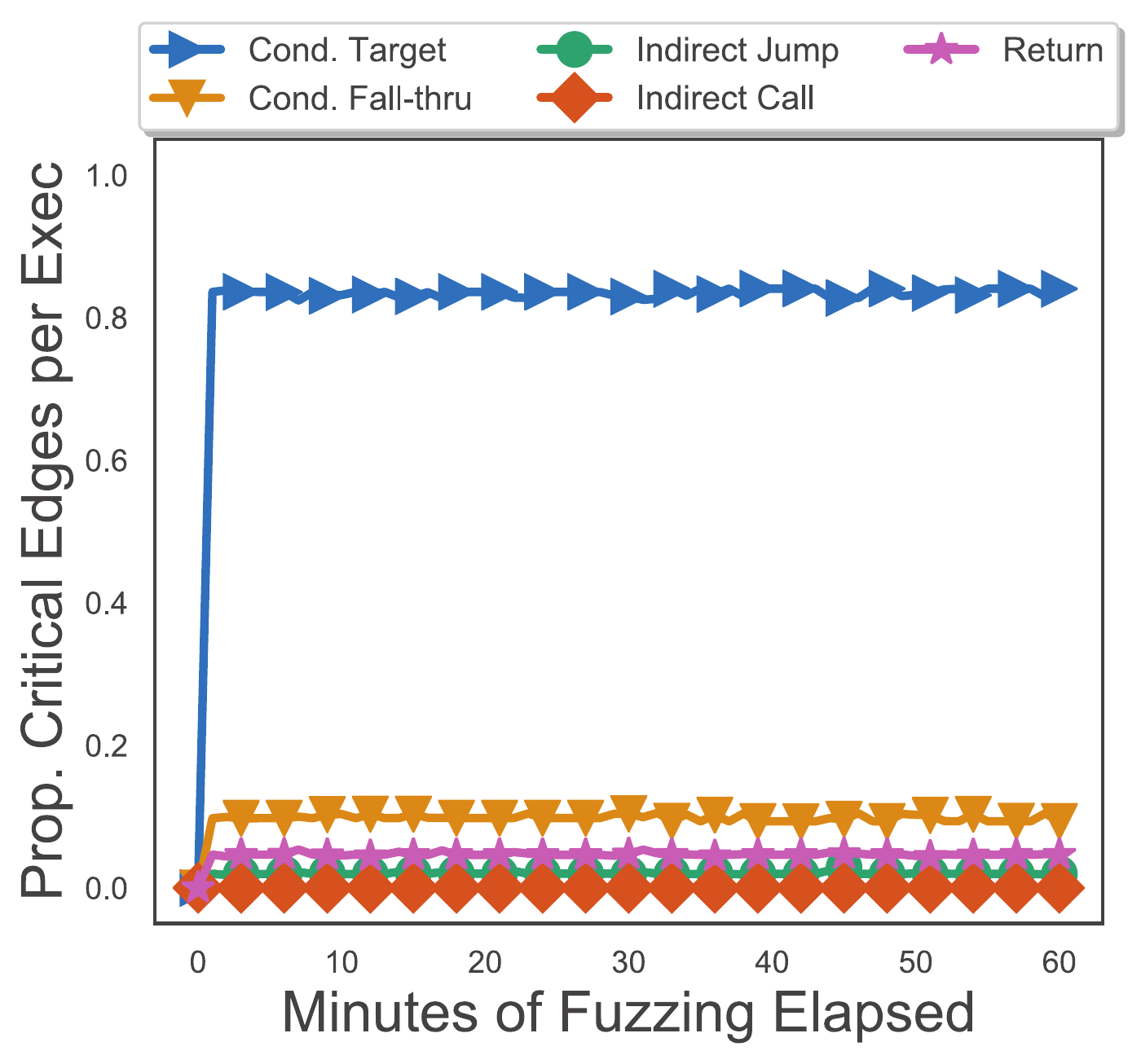}  
          \vspace{-0.45cm}
          \caption{\texttt{cert-basic}}
          \label{fig:critedgesdynamic:B}
     \end{subfigure}
     \begin{subfigure}[t]{0.236\textwidth}
          \includegraphics[width=\linewidth]{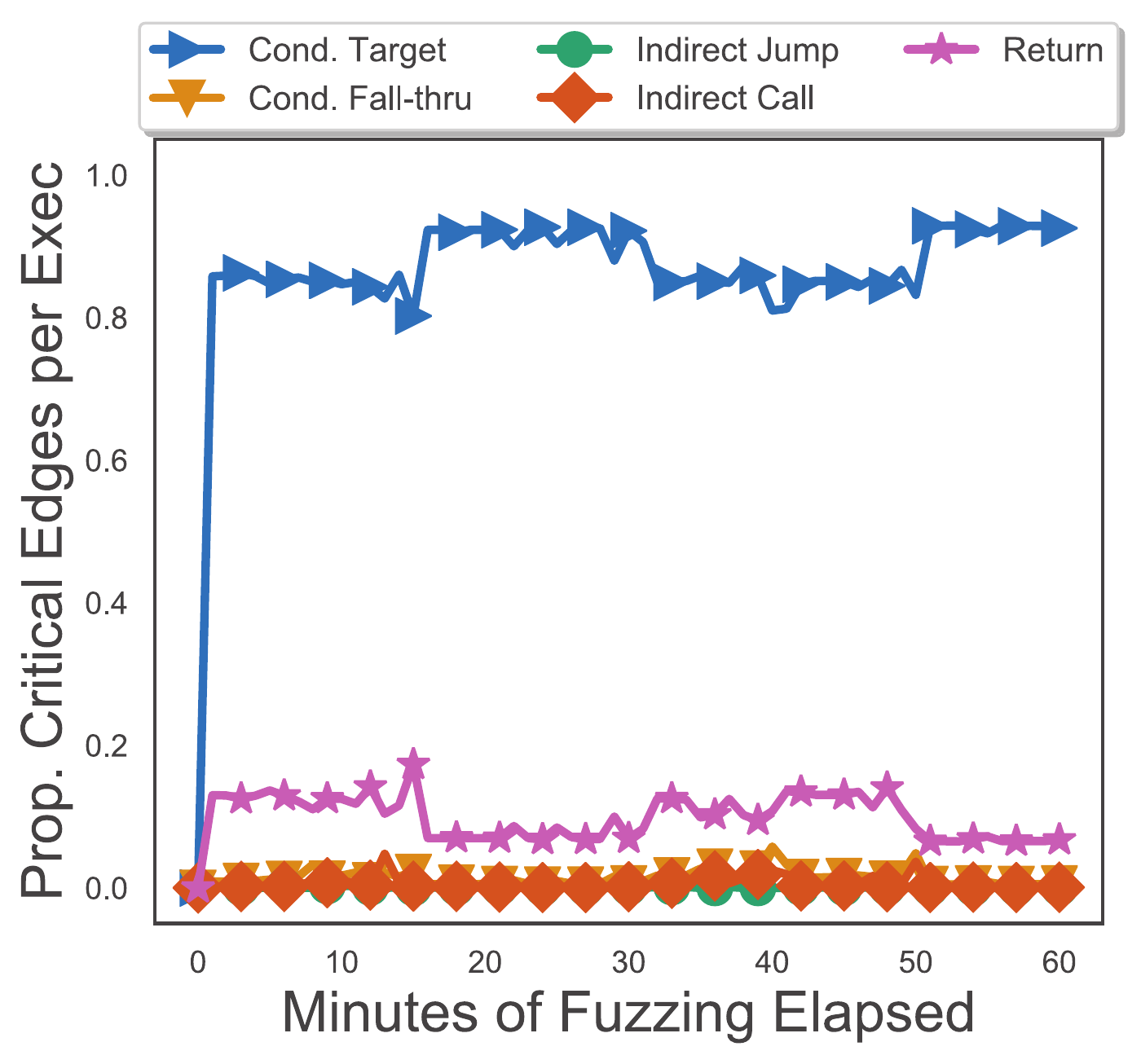}  
          \vspace{-0.45cm}
          \caption{\texttt{clean\_text}}
          \label{fig:critedgesdynamic:C}
     \end{subfigure}
     \begin{subfigure}[t]{0.236\textwidth}
          \includegraphics[width=\linewidth]{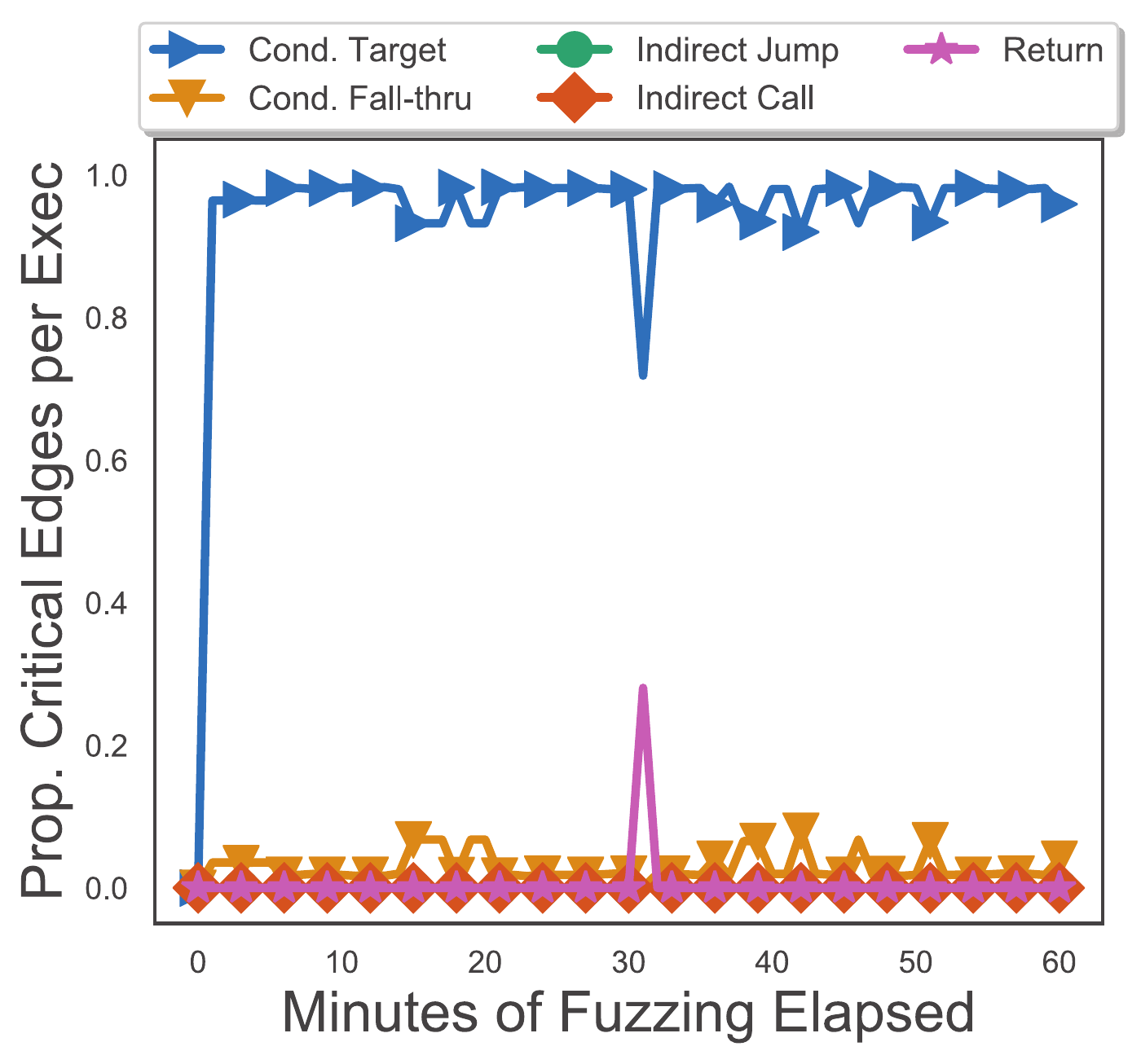}  
          \vspace{-0.45cm}
          \caption{\texttt{jasper}}
          \label{fig:critedgesdynamic:D}
     \end{subfigure}
    \vspace{-0.4cm}
    \caption{Visualization of the proportion of critical edges by transfer type encountered throughout fuzzing.}
    \label{fig:critedgesdynamic}
    \vspace{-0.1cm}
    \hrulefill
\end{figure}

\subsection{Supporting Edge Coverage}
\label{sec:covg:edgecovg}
AFL and its derivatives utilize \emph{hash-based} edge coverage, instrumenting each basic block to dynamically record edges as hashes of their start/end blocks.
However, as CGT's key speedup comes from replacing per-block instrumentation with far cheaper interrupts, it is thus incompatible with AFL-style hash-based edge coverage.

libFuzzer and honggFuzz  track edges using LLVM's SanitizerCoverage  instrumentation, which forgoes hashing to instead infer edges from the set of covered blocks.
For example, given a control-flow graph with edges $\vv{\texttt{ab}}$ 
and $\vv{\texttt{bc}}$, covering blocks \texttt{a} and \texttt{b} implies covering edge $\vv{\texttt{ab}}$; and subsequently covering \texttt{c} implies $\vv{\texttt{bc}}$.
However, such \emph{block-centric} edge coverage does not suffice if there exists a third edge $\vv{\texttt{ac}}$\eat{(\autoref{fig:critedges})}.
In this case, covering blocks \texttt{a}, \texttt{b}, and \texttt{c} implies edges $\vv{\texttt{ab}}$ and $\vv{\texttt{bc}}$; but since \texttt{c} has \emph{already} been covered, there is no way to detect $\vv{\texttt{ac}}$. 
Formally, these problematic edges are referred to as \emph{critical edges}: edges whose start/end blocks have two or more incoming/outgoing edges, respectively~\cite{the_clang_team_sanitizercoverage_2019}.

\begin{table}[!h]
\footnotesize    
    \centering
    
    \begin{tabular}[b]{p{1.5cm} | c | c : c }
        \specialrule{.1em}{0em}{0em} 
        \hline
        \textbf{Program} & \textbf{Total Edges} & \textbf{Crit. Edges} & \textbf{Prop.} \\ 
        \hline\hline
        \rowcolor{gray!10}\texttt{bsdtar} & 42911 & 9867 & 0.23 \\
        \texttt{cert-basic} & 7544 & 1642 & 0.22 \\
        \rowcolor{gray!10}\texttt{clean\_text} & 8762 & 1592 & 0.18 \\
        \texttt{jasper} & 21637 & 5878 & 0.27 \\
        \rowcolor{gray!10}\texttt{readelf} & 30959 & 7301 & 0.24 \\
        \texttt{sfconvert} & 8358 & 2022 & 0.24 \\
        \rowcolor{gray!10}\texttt{tcpdump} & 36200 & 7312 & 0.20 \\
        \texttt{unrtf} & 2505 & 465 & 0.19 \\
        \hline
        \multicolumn{3}{r}{\textbf{Mean}} & \textbf{22\%} \\
    \end{tabular}
    \vspace{-0.15cm}
    \caption{Proportion of critical edges in eight real-world programs.}
    \label{tab:critedgesstatic}
\vspace{-0.4cm}
\hrulefill{}
\vspace{0.2cm}
\end{table}
\begin{table}[!h]
    \footnotesize    
    \centering
    \vspace{-0.1cm}
    \begin{tabular}[b]{p{3.6cm} l }
        1. \textbf{Conditional target} & (e.g., \texttt{jle 0x100}'s \emph{True} branch) \\
        2. \textbf{Conditional fall-through} & (e.g., \texttt{jle 0x100}'s \emph{False} branch)  \\
        3. \textbf{Indirect jump} & (e.g., \texttt{jmp \%eax})  \\
        4. \textbf{Indirect call} & (e.g., \texttt{call \%eax})  \\
        5. \textbf{Return} & (e.g., \texttt{ret})  \\
    \end{tabular}
\vspace{-0.05cm}

\caption{Examples of x86 critical edge instructions by transfer type.}
\label{tab:critedgestypes}
\vspace{-0.4cm}
\hrulefill{}
\vspace{0.2cm}

\end{table}
\begin{table}[!ht]
\footnotesize
    
    \centering
    
    \begin{tabular}[b]{p{1.35cm} | c c c c c}
        \specialrule{.1em}{0em}{0em} 
        \hline
        \textbf{Program} & \textbf{CndTarg} & \textbf{CndFall} & \textbf{IndJmp} & \textbf{IndCall} & \textbf{Ret} \\
        \hline
        \hline
        \rowcolor{gray!10}\texttt{bsdtar} & 1.00 & 0.00 & 0.00 & 0.00 & 0.00 \\ 
        \texttt{cert-basic} & 0.84 & 0.10 & 0.02 & 0.00 & 0.05 \\
        \rowcolor{gray!10}\texttt{clean\_text} & 0.87 & 0.02 & 0.00 & 0.01 & 0.10 \\
        \texttt{jasper} & 0.97 & 0.03 & 0.00 & 0.00 & 0.00 \\
        \rowcolor{gray!10}\texttt{readelf} & 0.70 & 0.03 & 0.01 & 0.12 & 0.14 \\
        \texttt{sfconvert} & 0.84 & 0.02 & 0.00 & 0.00 & 0.13 \\
        \rowcolor{gray!10}\texttt{tcpdump} & 0.98\eat{0.40} & 0.01 & 0.00 & 0.00 & 0.01 \\
        \texttt{unrtf} & 0.94 & 0.03 & 0.00 & 0.00 & 0.02 \\
        \hline
        \multicolumn{1}{r}{\textbf{Mean}} & \textbf{89.3\%} & \textbf{2.9\%} & \textbf{0.4\%} & \textbf{1.6\%} & \textbf{5.7\%} \\
    \end{tabular}
    \vspace{-0.1cm}
    \caption{Proportion of encountered critical edges by transfer type.}
    \label{tab:critedgesdynamic}
\vspace{-0.4cm}
\hrulefill{}
\end{table}

\par{\textbf{Diving deeper into critical edges:}}
Supporting block-centric edge coverage requires resolving all critical edges.
LLVM's SanitizerCoverage achieves this by \emph{splitting} each critical edge with a ``dummy'' block, creating two new edges.
Continuing example \autoref{sec:covg:edgecovg}, dummy \texttt{d} will split critical edge $\vv{\texttt{ac}}$ into $\vv{\texttt{ad}}$ and $\vv{\texttt{dc}}$, thus permitting path $\vv{\texttt{a\textbf{d}c}}$ to be differentiated from $\vv{\texttt{a\textbf{b}c}}$.
But while such approach is indeed compatible with CGT's block-centric, interrupt-driven coverage, our analysis of eight real-world binaries shows \textbf{over~1~in~5} edges are critical (\autoref{tab:critedgesstatic}), revealing that splitting \emph{every} critical edge with a new block leaves a significant control-flow footprint---and inevitably, a higher baseline binary fuzzing overhead.

To understand the impact of critical edges on fuzzing, we instrument the same eight real-world binaries and dynamically record their instruction traces.\footnote{We limit instruction tracing to one hour of fuzzing due to the massive size of the resulting trace data (ranging from 200GB to 7TB per benchmark).}
In conjunction with the statically-generated control-flow graphs, we analyze each trace to measure the occurrences of critical edges; and further quantify them by \emph{transfer type}, which on the x86 ISA takes on one of five forms (shown in \autoref{tab:critedgestypes}).\footnote{As critical edges are, by definition, one of at least two outgoing edges from their starting block, transfers with at most one destination (direct jumps/calls and unconditional fall-throughs) can \emph{never} be critical edges.}

As shown in \autoref{fig:critedgesdynamic} and \autoref{tab:critedgesdynamic}, our findings reveal that \emph{conditional jump target} branches make up an average of \textbf{89\%} of all dynamically-encountered critical edges.

\vspace{0.1cm}
\stevebox{}{
\small
\textbf{Observation 1:} Conditional jump target branches make up the vast majority of critical edges encountered during fuzzing.
}

\subsubsection{\textbf{Jump Mistargeting}}
\label{sec:covg:jmpmis}
Splitting critical edges with dummy blocks adds a significant number of new instructions to each execution, and with it, more runtime overhead---slowing binary-only fuzzing down even further.
For the common case of critical edges (conditional jump target branches), we observe that the edge's destination address is encoded within the jump instruction itself, and thus can be statically altered to direct the edge elsewhere.
Our approach, \emph{jump mistargeting}, exploits this phenomena to ``mistarget'' the jump's destination so that it resolves into a CGT-style interrupt---permitting a signaling of the critical edge's coverage \emph{without} any need for a dummy block (i.e., identifying edge $\vv{\texttt{ac}}$ in \autoref{sec:covg:edgecovg}'s example\eat{\autoref{fig:critedges}} without the additional dummy block \texttt{d}).

\par{\textbf{An overview of jump addressing:}}
The x86 ISA has three types of jumps: \emph{short}, \emph{near} (or long), and \emph{far}.
Short and near jumps achieve intra-segment transfer via program counter (PC)-\emph{relative addressing}:
short jumps use 8-bit signed displacements, and thus can reach up to +127/-128 bytes relative to the PC;
while near jumps use much larger 16--32-bit signed displacements.
In contrast, far jumps achieve inter-segment transfer via \emph{absolute addressing} (i.e., to a fixed location irrespective of the PC).
All three jumps share the common instruction layout of an \emph{opcode} followed by a 1--4 byte \emph{destination operand} (an encoding of the relative/absolute address).
Since the adoption of position-independent layouts, most x86/x86-64 code utilizes relative addressing.

\par
\begin{figure}[!ht]
    \centering
    \includegraphics[width=\linewidth, trim=0 0 0 0 0]{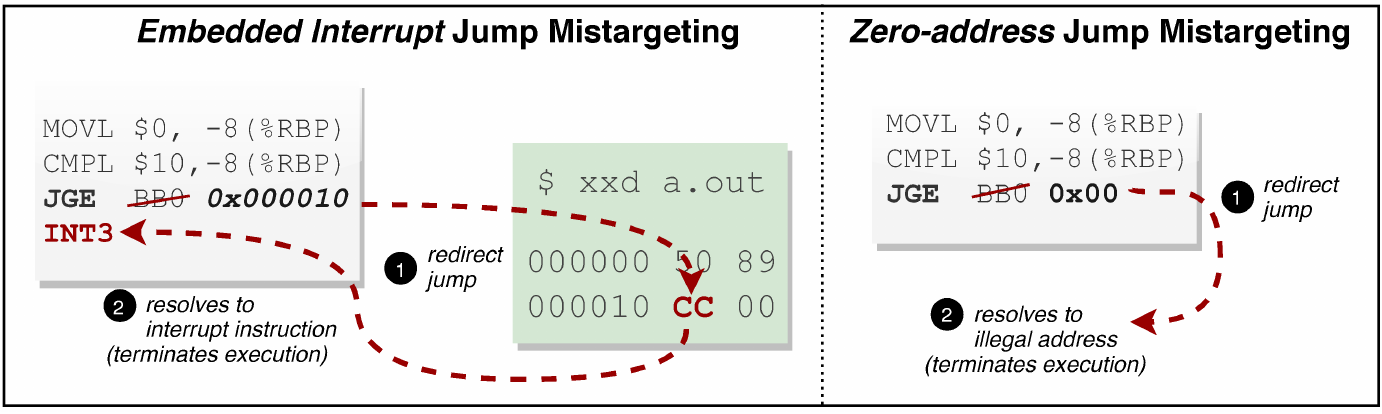}
    \vspace{-0.75cm}
    \caption{A visualization of jump mistargeting via embedded (left) and zero-address interrupts (right).}
    \label{fig:jmpmis}
    \vspace{-0.25cm}
    \hrulefill
\end{figure}

{\textbf{Redirecting jumps to interrupts:}}
Jump mistargeting alters conditional jump target critical edges to trigger interrupts when taken.
When used in CGT, its effect is identical to combining interrupts with conventional (yet more invasive) edge splitting---while avoiding the associated cost of inserting new blocks.
We envision two possible jump mistargeting strategies (\autoref{fig:jmpmis}): one leveraging \emph{embedded} interrupts, and another with \emph{zero-address} interrupts.

\begin{enumerate}
    \item \textbf{\emph{Embedded Interrupts}}.
    The simplest mistargeting approach is to replace each jump's destination with a garbage address, ideally resolving to an illegal instruction (thus interrupting the program).
    However, as many instructions have \emph{one-byte} opcodes, a carelessly-chosen destination may very well initiate an erroneous sequence of instructions.
    
    A more complete strategy is to instead redirect the jump to a location where an interrupt opcode is embedded.
    For example, the byte sequence \texttt{[00~CC]} at address \texttt{0x405500} normally resolves to instruction \texttt{[add~\%cl,\%ah]}; but as \texttt{0xCC} is itself an opcode for interrupt \texttt{int3}, it suffices to redirect the target critical edge jump to \texttt{0x40550\textbf{1}}, which subsequently fetches and executes \texttt{0xCC}, thus triggering the interrupt instruction. 
    A key challenge (and bottleneck) of this approach is scanning the bytespace in the jump's displacement range to pinpoint embedded interrupts.
    
    \item \textbf{\emph{Zero-address Interrupts}}.
    As nearly all x86/x86-64 code is position-independent and hence uses PC-relative addressing, an alternative and less analysis-intensive mistargeting approach is to interrupt the program by resolving the jump's displacement to the zero address (i.e., \texttt{0x00}).
    For example, taking the conditional jump represented by byte sequence \texttt{[0F~8F~7C~00~00~00]} at address \texttt{0x400400} normally branches to address \texttt{0x400400}+\texttt{6}+\texttt{0x0000007C} (i.e., the PC + instruction length + displacement); but to resolve it to the zero address merely requires the displacement be rewritten to \texttt{0xFFBFFB7E} (i.e., the negative sum of the PC and instruction length).
    As 8--16 bit displacements do not provide enough ``room'' to cover the large virtual address space of modern programs, zero-address mistargeting is generally restricted to jumps with 32-bit displacements, however, most x86-64 branches fit this mold.
\end{enumerate}

\stevebox{}{
\small
\textbf{Technique 1:} Jumps' self-encoded targets can be rewritten to resolve to addresses that result in interrupts, enabling binary-level CGT edge coverage at native speed (i.e., without needing to insert additional basic blocks).
}

\subsection{Supporting Hit Counts}
Most fuzzers today adopt AFL-style~\cite{zalewski_american_2017} bucketed hit count coverage, which coarsely tracks changes in block/edge execution frequencies over a set of eight ranges: 0--1, 2, 3, 4--7, 8--15, 16--31, 32--127, and 128+.
Unfortunately, CGT's interrupt-driven coverage currently only supports a \emph{binarized} notion of coverage (i.e., taken/not taken), and thus requires a fundamentally new approach to support finer-grained frequencies. 

\par{\textbf{Diving deeper into hit counts:}}
In exploring the importance of hit counts, we observe that most new hit count coverage is localized to \emph{loops} (e.g., \texttt{for()}, \texttt{while()}).
As Rawat and Mounier~\cite{rawat_vuzzer_2017} demonstrate that as many as \textbf{42\%} of binary code loops induce buffer overflows (e.g., by iterating over user-provided input with \texttt{strcpy()}), it is imperative to track hit counts as a means of assessing---and prioritizing---fuzzer ``progress'' toward higher loop iterations.
However, inferring a loop's iteration count is achievable purely from monitoring its induction variable---eliminating the expense of tracking hit counts for \emph{every} loop block (as AFL and libFuzzer do).

\vspace{0.1cm}
\stevebox{}{
\small
\textbf{Observation 2:} 
Hit counts provide fuzzing a notion of loop exploration progress, but need only be tracked once per loop iteration.
}

\subsubsection{\textbf{Bucketed Unrolling}}
AFL-style~\cite{zalewski_american_2017} hit count tracking adds counters to each block/edge to dynamically update their respective hit counts in a shared memory coverage bitmap.
However, this approach is fundamentally incompatible with the binarized nature of CGT's block-centric, interrupt-driven coverage.
While a naive solution is to instead add CGT's interrupts following the application of a \emph{loop peeling} transformation---making several copies of the loop's body and stitching them together with direct jumps (e.g., \texttt{head}~$\rightarrow$~{\texttt{body}$_1$}~$\rightarrow$~...~$\rightarrow$~{\texttt{body}$_n$} $\rightarrow$~\texttt{tail})---the resulting binary will be exceedingly space inefficient due to excessive code duplication---especially for nested loops.

In search of a more performant solution, we develop \emph{bucketed unrolling}---drawing from compiler loop unrolling principles to encode the functionality of AFL-style bucketed hit counts as a series of binarized range comparisons.

\begin{figure}[h]
    \centering
    \includegraphics[width=0.95\linewidth]{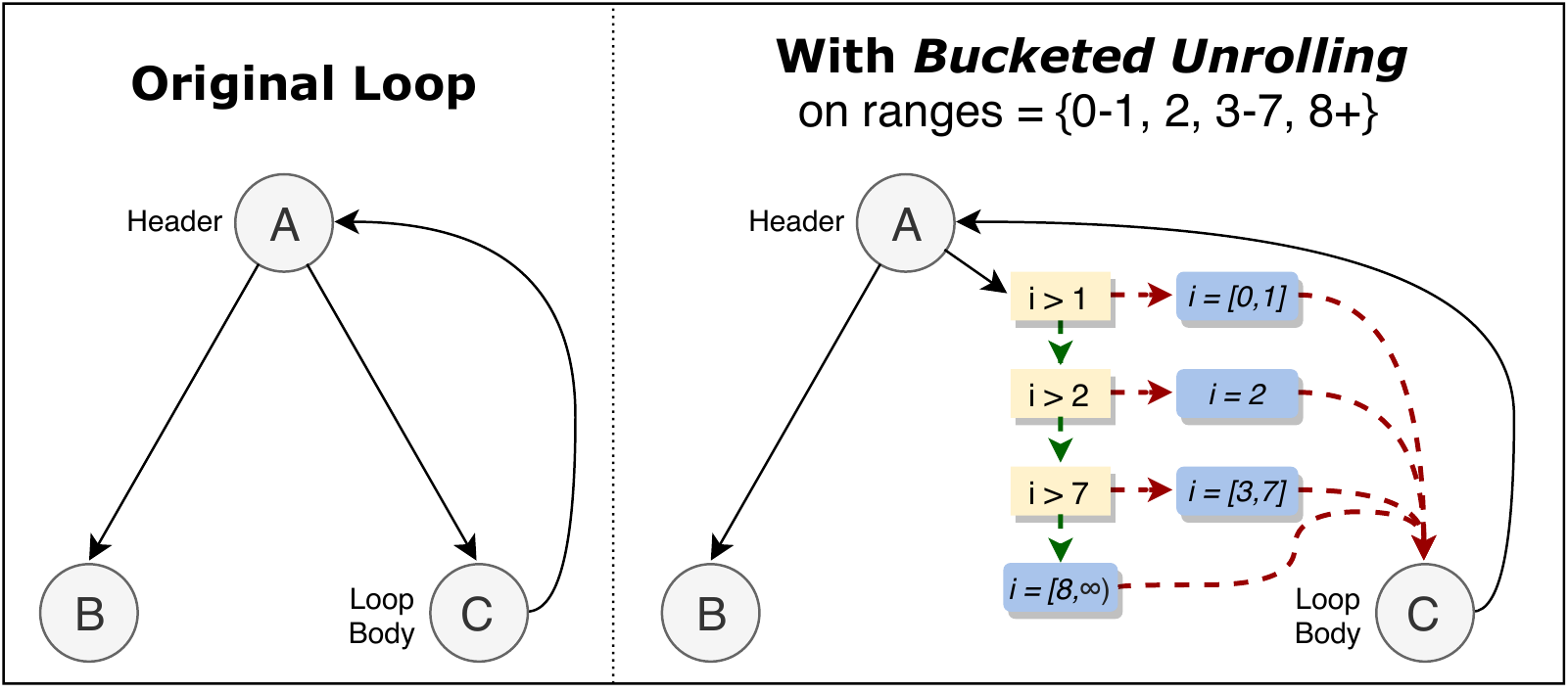}
    \vspace{-0.375cm}
    \caption{Bucketed unrolling applied to a simple loop.}
    \label{fig:bunroll}
    \vspace{-0.25cm}
    \hrulefill
\end{figure}

As shown in \autoref{fig:bunroll}, bucketed unrolling augments each loop header with a series of sequential conditional statements weighing the loop induction variable against the desired hit count bucket ranges (e.g., AFL's eight).
To support CGT, each conditional's fall-through block is assigned an interrupt;
taking any conditional's target branch jumps directly to the loop's body, indicating no change from the current bucket range; and taking the fall-through triggers the next sequential interrupt, thus signaling an advancement to the next bucket.
The resulting code replicates the functionality of AFL-style hit count tracking---but obtains much higher performance by doing so at just one instrumentation location per loop.

\vspace{0.1cm}
\stevebox{}{
\small
\textbf{Technique 2:} 
Encoding conventional bucketed hit count tracking as a series of sequential, binarized range checks enables CGT to capture binary-level loop exploration progress---while upholding its fast, interrupt-driven coverage-tracing strategy.
}

\section{Implementation: \platname}
In this section we introduce \platname---\emph{\textbf{H}igh-\textbf{E}fficiency e\textbf{X}panded \textbf{C}overage for \textbf{I}mproved \textbf{T}esting of \textbf{E}xecutables}---our implementation of binary-only \emph{coverage-preserving} Coverage-guided Tracing.
Below we discuss \platname's core architecture, and our design decisions in realizing jump mistargeting and bucketed unrolling.

\subsection{Architectural Overview} 
\platname consists of three main components: \textbf{(1)~binary generation}, \textbf{(2)~control-flow mapping}, and \textbf{(3)~the fuzzer}.
We implement components 1--2 as a set of analysis and transformation passes atop the ZAFL static rewriting platform~\cite{nagy_breaking_2021}, and component 3 atop the industry-standard fuzzer AFL~\cite{zalewski_american_2017}. 
Below we briefly discuss each and their synergy in facilitating coverage-preserving CGT.

\par{\textbf{Binary Generation:}}
\platname's workflow is similar in nature to UnTracer's~(\autoref{fig:cgt}); i.e., we generate two versions of the original target binary: (1) an \emph{oracle} (run for every test case) with interrupts added to each basic block; and (2) a \emph{tracer} (run only for coverage-increasing test cases) equipped with conventional tracing instrumentation.
While many fuzzers embrace compiler instrumentation for its speed and soundness (i.e., LLVM~\cite{lattner_llvm_2004}), there are by now a number of static binary rewriters with comparable qualities.
We examine several popular and/or emerging security-oriented binary rewriters---Dyninst~\cite{paradyn_tools_project_dyninst_2018}, McSema~\cite{dinaburg_mcsema_2014}, RetroWrite~\cite{dinesh_retrowrite_2020}, and ZAFL~\cite{nagy_breaking_2021}---and distill a set of properties we feel are best-suited supporting jump mistargeting and bucketed unrolling: (1) \emph{a modifiable control-flow} representation; (2) \emph{dominator flow analysis}~\cite{agrawal_dominators_1994}); and (3) \emph{sound code transformation and generation}.
We select ZAFL as the basis for \platname as it is the highest performance rewriter that possesses the above three properties in addition to an LLVM-like transformation API.
We expect that with additional engineering effort, our findings apply to the other rewriters listed.

Like most static binary rewriters, ZAFL operates by first disassembling and lifting the input binary to an intermediate representation;\footnote{ZAFL's disassembly supports mixing-and-matching of recursive descent and linear sweep. The current tools utilized are based on IDA Pro~\cite{guilfanov_ida_2019} and GNU objdump~\cite{gnu_project_gnu_2018}.}
and performing all code transformation at this IR level (e.g., injecting bucketed unrolling's range checks~\autoref{sec:impl:bunroll}), adjusting the binary's layout as necessary before reconstituting the final executable. 
While relocating direct (i.e., absolute and PC-relative) control flow is generally trivial, attempting so for \emph{indirect} transfers is \emph{undecidable} and risks corrupting the resulting binary, as their respective targets cannot be identified with any generalizable accuracy~\cite{wenzl_hack_2019, pang_sok_2021}.
ZAFL addresses this challenge conservatively via \emph{address pinning}~\cite{hiser_zipr_2017,hawkins_zipr_2017}, which ``pins'' any \emph{unmovable} items (including but not limited to: indirectly-called function entries, callee-to-caller return targets, data, or items that cannot be precisely disambiguated as being either code or data) to their \emph{original} addresses;\footnote{To support address pinning, ZAFL conservatively scans for addresses \emph{likely} targeted by indirect control flow; generally this is achieved via rudimentary heuristics (e.g., post-call instructions, jump table entries, etc.). Additionally, ZAFL pins all data items.} while safely relocating the remaining \emph{movable} items around these pins (often via chained jumps).
Though address pinning will likely over-approximate the set of unmovable items at slight cost to binary performance and/or space efficiency (particularly for exceedingly-complex binaries with an abundance of jump tables, handwritten assembly, or data-in-code), its \emph{general-purpose soundness, speed, and scalability}~\cite{nagy_breaking_2021} \emph{makes it promising for facilitating coverage-preserving CGT.}
Our current prototype, \platname, supports binary fuzzing of x86-64 Linux C and C++ executables.

\par{\textbf{Control-flow Mapping:}}
A key requirement of CGT is a mapping of each oracle basic block's address (i.e., where an interrupt is added) to its corresponding tracer binary trace-block ID;
when a coverage-increasing test case is found, the tracer is invoked to capture the test case's full coverage, for which all interrupts are subsequently removed at their addresses in the oracle.
To generate this mapping, we save the original and rewritten control-flow graphs for both the oracle and tracer binaries.
We then parse the pair of original control-flow graphs to find their corresponding matches, and subsequently map each to their oracle and tracer binary counterparts (i.e., \texttt{(cfgBB,oracleBB)} $\rightarrow$ \texttt{(cfgBB,tracerBB)}).
From there, we generate the necessary \texttt{(oracleAddr,tracerID,interruptBytes)} mapping for each block (e.g., \texttt{(0x400400,30,0xCC)}).
If mapping should fail (e.g., a tracer block with no corresponding oracle block), we omit the block to avoid problematic interrupts; we observe this generally amounts to no more than a handful of instances per binary, and does not impact \platname's overall coverage~(\autoref{sec:eval:covg:edge}--\autoref{sec:eval:covg:loop}).

\par
\begin{figure}[h!]
    \centering
    \framebox{\includegraphics[width=0.98\linewidth, trim={0.25ex 0.25ex 0.45ex 1.15ex},clip]{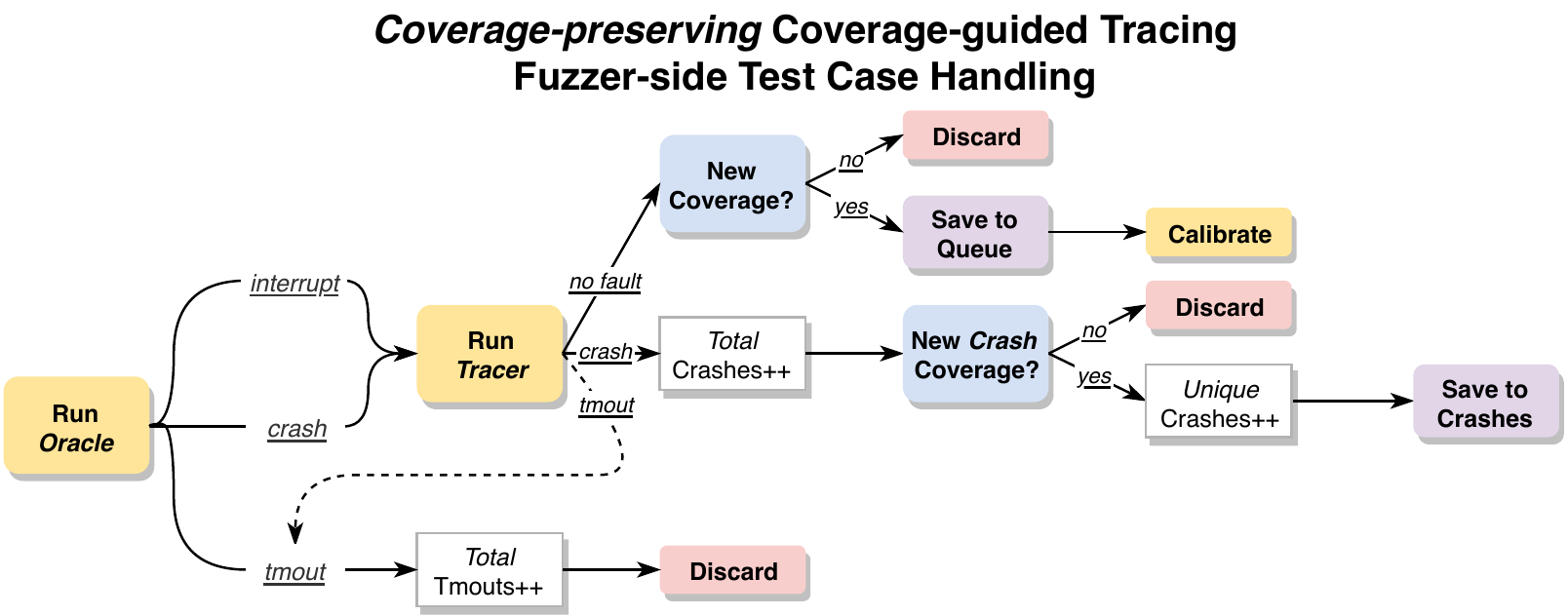}}
    \vspace{-0.575cm}
    \caption{\platname's fuzzer-side test case handling logic. Like UnTracer, we discard timeout-producing test cases; however, we re-run crashing test cases to determine whether they are a \emph{true} crash (i.e., occurring on both the oracle and tracer) or the result of hitting an oracle \emph{mistargeted} edge (generally triggering a \texttt{SIGSEGV} from the jump being redirected to the zero address). }
    \label{fig:zuntracer}
    \vspace{-0.26cm}
    \hrulefill
\end{figure}

{\textbf{The Fuzzer:}}
Like UnTracer, we implement \platname atop the industry-standard fuzzer AFL~\cite{zalewski_american_2017} 2.52b with several changes in test case handling logic~(\autoref{fig:zuntracer}).
We default to conventional tracing for any executions where coverage is required (e.g., calibration and trimming), while not re-executing or saving timeout-producing test cases.
As jump mistargeting triggers signals that might otherwise appear as valid crashes (e.g., \texttt{SIGSEGV}), we alter \platname's fuzzer-side crash-handling logic as follows: if a test case crashes the oracle, we re-run it on the tracer to verify whether it is a \emph{true} or a \emph{mistargeted} crash; if it does not crash the tracer, we conclude it is the result of taking a mistargeted critical edge (i.e., a \texttt{SIGSEGV} from jumping to the zero address), and save it to the fuzzer queue.
We note that the core principles of coverage-preserving CGT scale to any fuzzer (e.g., honggFuzz), as evidenced by emerging CGT-based efforts within the fuzzing community~\cite{gros_fuzzing_2020, jung_winnie_2021, fioraldi_afl_2020}.

\subsection{Implementing Jump Mistargeting} 
\label{sec:impl:jmpmis}
We implement \emph{zero-address} jump mistargeting for the common-case of critical edges, conditional jump target branches~(\autoref{sec:covg:edgecovg}), as follows. 
To statically identify critical edges we first enumerate all control-flow edges, and mark an edge as \emph{critical} if at least two edges both precede and succeed it.
We subsequently parse each critical edge and categorize it by type by examining its starting block's last instruction~(\autoref{tab:critedgestypes}).
Lastly, we update an offline record of each critical edge by type (e.g., ``conditional~jump~target'') and its respective starting/ending basic block addresses.

We enumerate all conditional jump target critical edges;
as x86-64 conditional jumps are 6-bytes in length and encoded with a 32-bit PC-relative displacement, we compute the sum of the instruction's address and its length, and determine the 2's complement (i.e., negative binary representation).
Using basic file I/O we then statically overwrite the jump's displacement operand with the little-endian encoding of the zero-address-mistargeted displacement, and update our oracle-to-tracer mapping accordingly (e.g., \texttt{(0x400400,30,0x7C000000)} for the example in \autoref{sec:covg:edgecovg}).

If a critical edge cannot accommodate zero-address mistargeting (e.g., from having a $<$32-bit displacement), we attempt to fall-back to conventional SanitizerCoverage-style~\cite{the_clang_team_sanitizercoverage_2019} edge splitting, inserting a dummy block and connecting it to the edge's end block.
Conditional \emph{fall-through} critical edges require careful handling, as accommodating the transfer from the edge's starting block to the dummy requires the dummy be placed immediately after the starting block (i.e., the next sequential address).
However, splitting \emph{indirect} critical edges remains a universal problem even for robust compilers like LLVM~(\autoref{sec:disc:indy}). 
While recent work~\cite{kim_refining_2021} reveals the possibility that indirect edges may be modeled at the binary level, such approaches are still too imprecise to be realistically deployed; hence, we conservatively omit indirect critical edges as we observe they have little overall significance on dynamically-seen control-flow~(\autoref{fig:critedgesdynamic}).

\subsection{Implementing Bucketed Unrolling}
\label{sec:impl:bunroll}
We implement bucketed unrolling to replicate AFL-style loop hit count tracking, beginning with an analysis pass to retrieve all code loops from the target binary based on the classic \emph{dominance-based} loop detection~\cite{ramalingam_loops_2002}: given the control-flow graph and dominator tree (generally available in any off-the-shelf static rewriter's API), we mark a set of blocks \texttt{S} as a loop if (1) there exists a header block \texttt{h} that dominates all blocks in \texttt{S}; and (2) there exists a \emph{backward} edge $\vv{\texttt{b}\texttt{h}}$ from some block $\texttt{b}\in \texttt{S}$ such that \texttt{h} dominates \texttt{b}.\footnote{In compiler and graph theory, a basic block \texttt{a} is said to \emph{dominate} basic block \texttt{b} if and only if every path through \texttt{b} also covers \texttt{a}.~\cite{agrawal_dominators_1994}}
Though binary-level loop head/body detection is difficult---particularly around complex optimizations like Loop-invariant Code Motion---we observe that the standard dominance-based algorithm is sufficient; and while \platname attains the highest loop coverage in our evaluation~(\autoref{sec:eval:covg:loop}), we expect that future advances in optimized-binary loop detection will only improve these capabilities.

As pinpointing a loop's induction variable (the target of bucketed unrolling's discrete range checks) is itself semantically challenging at the binary level, we opt for a simpler approach and instead add a ``fake'' loop counter before each loop header; and augment the header with an instruction to increment this counter per iteration (e.g., x86's \texttt{incl}).
\emph{Where} the increment is inserted in the header ultimately depends on the static rewriter of choice;
Dyninst~\cite{paradyn_tools_project_dyninst_2018} prefers to conservatively insert new code at basic block entrypoints to avoid clobbering occupied registers; while RetroWrite~\cite{dinesh_retrowrite_2020} and ZAFL~\cite{nagy_breaking_2021} analyze register liveness to more tightly weave code with the original instructions.
Either style is supportive of \platname, though tight code insertion is preferable for higher runtime speed.

We implement bucketed unrolling's sequential range checks (per AFL's 8-bucket hit count scheme) as a transformation pass directly before the loop's first body block; and connect each to the first body block via direct jumps, and to each other via fall-throughs. The resulting assembly resembles the following (shown in Intel syntax):

\eat{
\begin{stevemint}
_loop_head:
    incl  rdx
    cmpl  1, rdx
    jle    _loop_body
    cmpl  2, rdx
    jle    _loop_body
    ...
_loop_body:
\end{stevemint}
}

\begin{stevemint}
_loop_head:
    incl  rdx
    cmpl  rdx, 1
    jle    _loop_body
    cmpl  rdx, 2
    jle    _loop_body
    ...
_loop_body:
\end{stevemint}
\vspace{-0.1cm}

To facilitate signaling of a range change, we flag the start of each sequential range check (e.g., lines 3 and 5 above) with the one-byte \texttt{0xCC} interrupt.
To maintain control-flow congruence, we apply this transformation to both the oracle and tracer binaries.

\section{Evaluation}
Our evaluation of the effectiveness of \emph{coverage-preserving Coverage-guided Tracing} is motivated by three key questions:

\begin{itemize}
    \setlength\itemsep{-0.0em}
    \item[\textbf{Q1:}] Do jump mistargeting and bucketed unrolling improve coverage over basic-block-only CGT? 
    \item[\textbf{Q2:}] What are the performance impacts of expanding CGT to finer-grained code coverage metrics? 
    \item[\textbf{Q3:}] How do the benefits of coverage-preserving CGT impact fuzzing bug-finding effectiveness?
\end{itemize}

\subsection{Experiment Setup}
Below we provide expanded detail on our evaluation: the coverage-tracing approaches we are testing, our benchmark selection, and our experimental infrastructure and analysis procedures.

\par\textbf{Competing Tracing Approaches:}
\autoref{tab:configs} lists the fuzzing coverage-tracing approaches tested in our evaluation.
We evaluate our binary-only coverage-preserving CGT implementation, \platname, alongside the current \emph{block-coverage-only} CGT approach \textbf{UnTracer}~\cite{nagy_full-speed_2019}.\footnote{As UnTracer is partially reliant on AFL's \emph{source-level} instrumentation and is hence impossible to use on binary-only targets in its original form, we implement a \emph{fully binary-only} version suitable across all 12 of our evaluation benchmarks.}
To test \platname's fidelity against the conventional \emph{always-on} coverage tracing in binary fuzzing, we also evaluate the leading binary tracers \textbf{QEMU} (AFL~\cite{zalewski_american_2017} and honggFuzz's~\cite{swiecki_honggfuzz_2018} default approach for fuzzing binary-only targets); \textbf{Dyninst} (a popular static-rewriting-based alternative~\cite{heuse_afl-dyninst_2018}); and \textbf{RetroWrite}~\cite{dinesh_retrowrite_2020} (a recent static-rewriting-based instrumenter).
Lastly, we replicate UnTracer's evaluation for open-source targets by further comparing against \textbf{AFL-Clang} (AFL's~\cite{zalewski_american_2017} source-level always-on tracing)~\cite{nagy_full-speed_2019}. 
We report \platname's best-performing coverage configuration (edge coverage or edge+count coverage) in all experiments.

\begin{table}[h!]
\footnotesize 
    \centering
    
    \begin{tabular}[b]{p{1.9cm} | c | c | c }
        \specialrule{.1em}{0em}{0em} 
        \hline
        \textbf{Approach} & \textbf{Tracing Type} & \textbf{Level } &  \textbf{Coverage} \\ 
         \hline\hline
        \rowcolor{orange!15}\texttt{HeXcite} & coverage-guided & binary & edge + counts \\
        
        \texttt{UnTracer}~\cite{nagy_full-speed_2019} & coverage-guided & binary & block \\  
        
        \hdashline
        
        \rowcolor{white}\texttt{QEMU}~\cite{zalewski_american_2017} & always-on & binary & edge + counts \\
        
        \texttt{Dyninst}~\cite{heuse_afl-dyninst_2018} & always-on & binary & edge + counts \\
        
        \rowcolor{white}\texttt{RetroWrite}~\cite{dinesh_retrowrite_2020} & always-on & binary & edge + counts \\
        
        \hdashline
        
        \texttt{Clang}~\cite{zalewski_american_2017} & always-on & source & edge + counts \\

    \end{tabular}
    \vspace{-0.00cm}
    \caption{Fuzzing coverage tracers evaluated alongside \platname; and their \emph{type}, \emph{level}, and \emph{coverage} metric.}
    \label{tab:configs}
\vspace{-0.4cm}
\hrulefill{}
\end{table}

\begin{table}[!h]
\footnotesize 
    \centering
    
    \vspace{-0.5cm}
    
    \begin{tabular}[b]{p{1.5cm} | c | c | c }
        \specialrule{.1em}{0em}{0em} 
        \hline
        \textbf{Binary} & \textbf{Package} & \textbf{Source} & \textbf{Input File} \\ 
        \hline\hline    
        \rowcolor{gray!10}\texttt{jasper} & jasper-1.701.0 & \cmark & JPG \\
        \texttt{mjs} & mjs-1.20.1 & \cmark & JS  \\
        \rowcolor{gray!10}\texttt{nasm} & nasm-2.10 & \cmark & ASM  \\
        \texttt{sam2p} & sam2p-0.49.3 & \cmark & BMP  \\
        \rowcolor{gray!10}\texttt{sfconvert} & audiofile-0.2.7 & \cmark & WAV  \\
        \texttt{tcpdump} & tcpdump-4.5.1 & \cmark & PCAP  \\
        \rowcolor{gray!10}\texttt{unrtf} & unrtf-0.20.0 & \cmark & RTF  \\
        \texttt{yara} & yara-3.2.0 & \cmark & YAR  \\
        \hline
        \rowcolor{gray!10}\texttt{lzturbo} & lzturbo-1.2 & \xmark & LZT \\
        \texttt{pngout} & Mar 19 2015 & \xmark & PNG \\
        \rowcolor{gray!10}\texttt{rar} & rarlinux-4.0.0 & \xmark & RAR \\
        \texttt{unrar} & rarlinux-4.0.0 & \xmark & RAR \\
    \end{tabular}
    \vspace{-0.1cm}
    \caption{Our evaluation benchmark corpora.}
    \label{tab:benchmarks}
    \vspace{-0.4cm}
    \hrulefill
    \vspace{0.2cm}

\end{table}

\par\textbf{Benchmark Selection:}
Our benchmark selection (\autoref{tab:benchmarks}) follows the current standard in the fuzzing literature, consisting of eight binaries from popular open-source applications varying by input file format (e.g., images, audio, video) and characteristics.
Furthermore, as CGT's most popular usage to date~\cite{gros_fuzzing_2020, jung_winnie_2021, fioraldi_afl_2020} is in accelerating \emph{binary-only} fuzzing, we also incorporate a set of four closed-source binary benchmarks distributed as free software.
All benchmarks are selected from versions with well-known bugs to ensure a \emph{self-evident} comparison in our bug-finding evaluation.

For each tracing approach we omit benchmarks that are unsupported or fail: \texttt{sam2p} and \texttt{sfconvert} for QEMU (due to repeated deadlock); \texttt{lzturbo}, \texttt{pngout}, \texttt{rar}, and \texttt{unrar} for Dyninst (due to its inability to support closed-source, stripped binaries~\cite{nagy_breaking_2021}); \texttt{jasper}, \texttt{nasm}, \texttt{sam2p}, \texttt{lzturbo}, \texttt{pngout}, \texttt{rar}, and \texttt{unrar} for RetroWrite (due to crashes on startup and/or being position-dependent/stripped); and \texttt{lzturbo}, \texttt{pngout}, \texttt{rar}, and \texttt{unrar} for AFL-Clang (due to it only supporting open-source targets).


\begin{figure*}[t!]
    \centering
    \tiny
     \begin{subfigure}[t]{0.1975\textwidth}
          \includegraphics[width=\linewidth, trim=5 0 4.5 0]{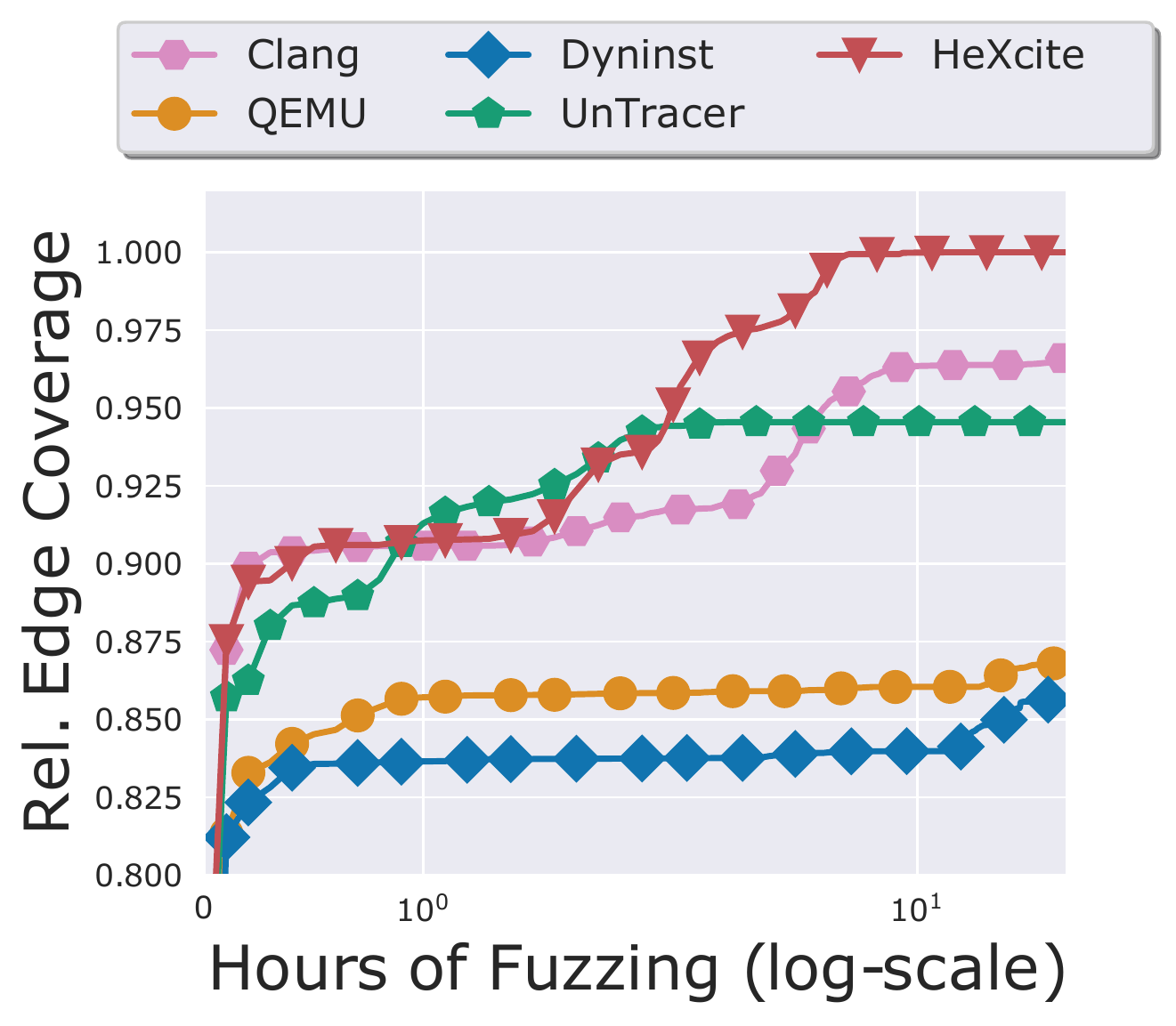}
          \vspace{-0.4cm}
          \caption{ \texttt{nasm}}
          \label{fig:coverage:edge:A}
     \end{subfigure}
     \begin{subfigure}[t]{0.1975\textwidth}
          \includegraphics[width=\linewidth, trim=5 0 4.5 0]{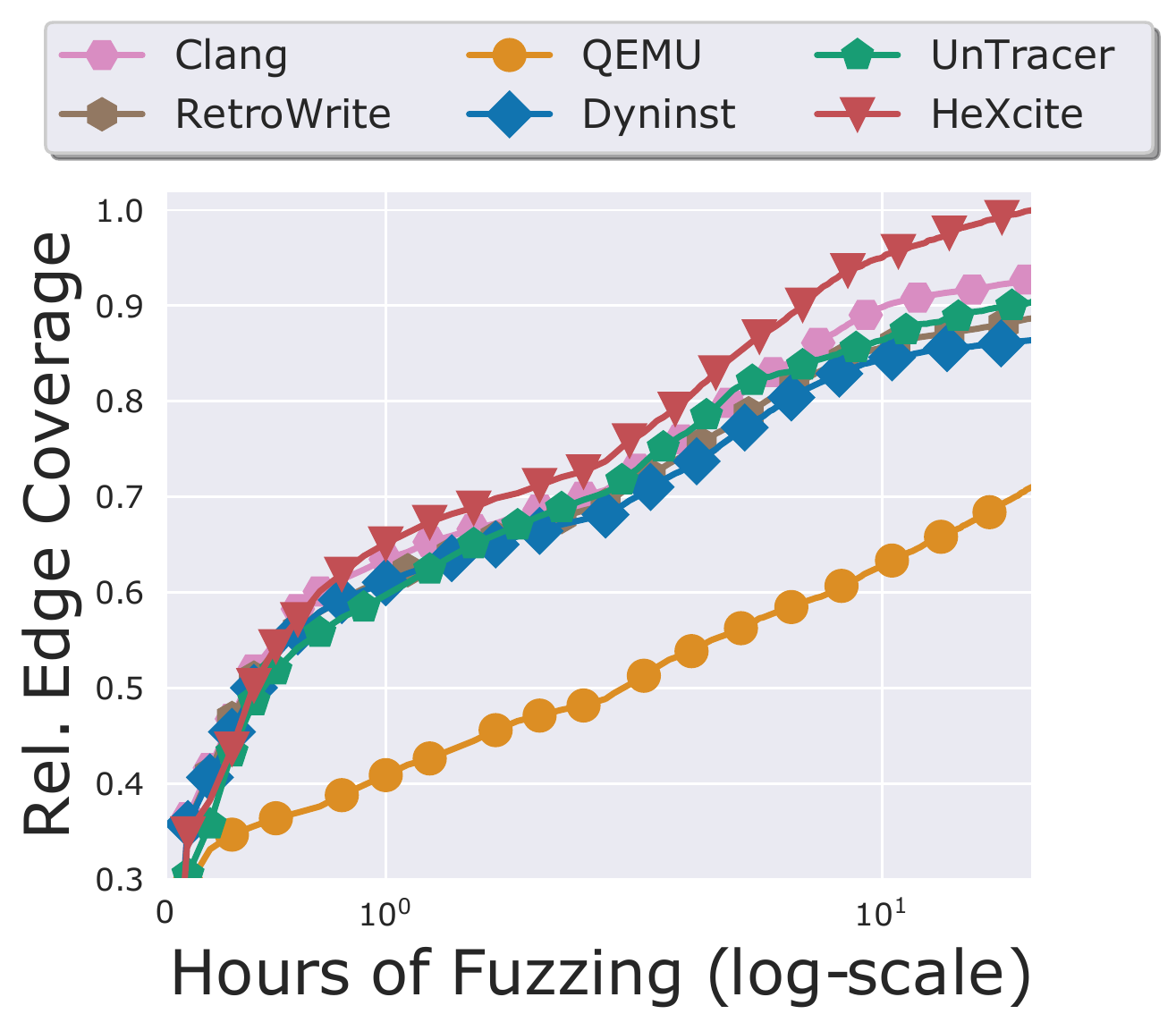}  
          \vspace{-0.4cm}
          \caption{\texttt{tcpdump}}
          \label{fig:coverage:edge:B}
     \end{subfigure}
     \begin{subfigure}[t]{0.2\textwidth}
          \includegraphics[width=\linewidth, trim=5 0 4.5 0]{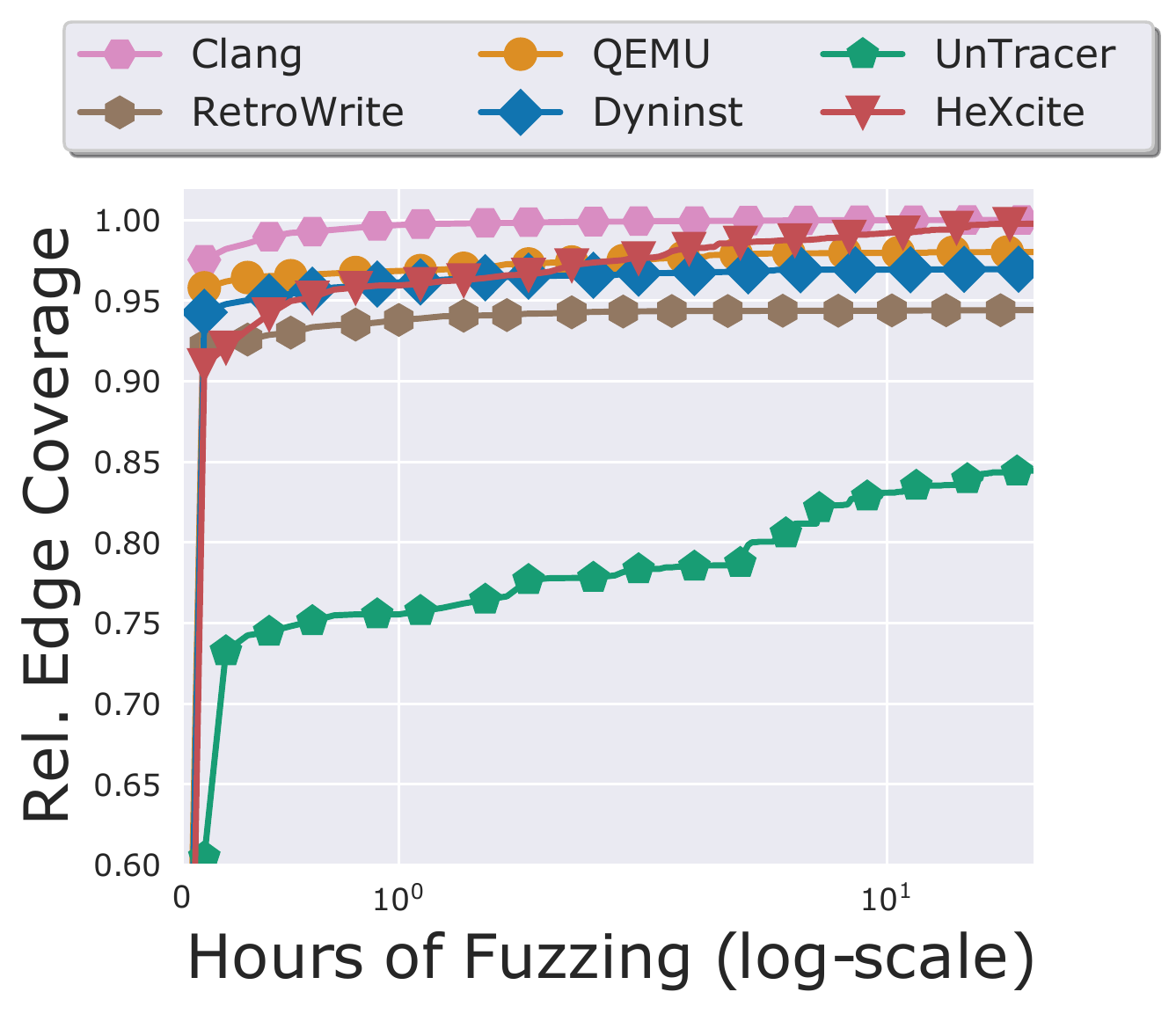}  
          \vspace{-0.4cm}
          \caption{\texttt{unrtf}}
          \label{fig:coverage:edge:C}
     \end{subfigure}
     \begin{subfigure}[t]{0.1975\textwidth}
          \includegraphics[width=\linewidth, trim=5 0 4.5 0]{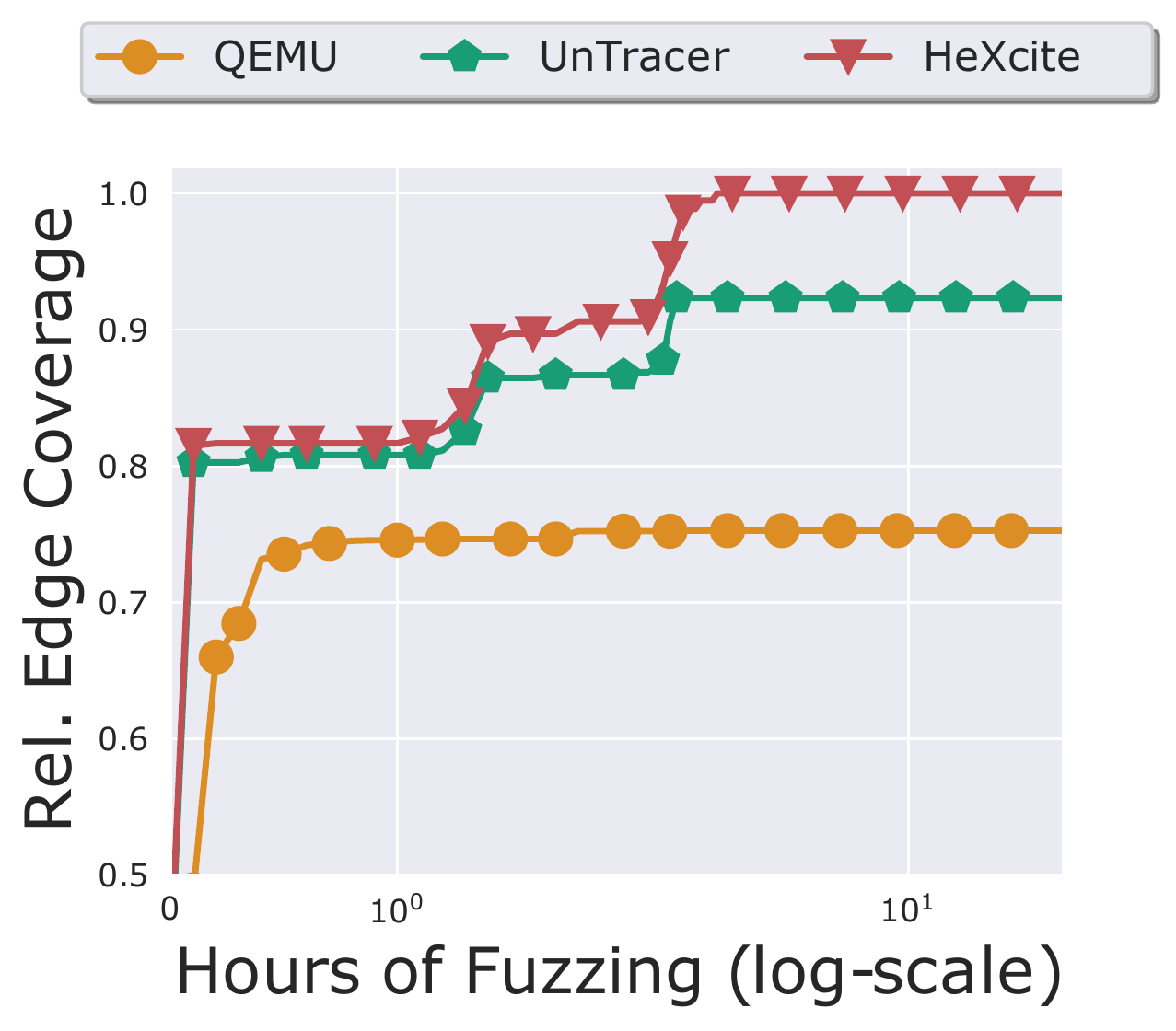}  
          \vspace{-0.4cm}
          \caption{\texttt{pngout}}
          \label{fig:coverage:edge:D}
     \end{subfigure}
     \begin{subfigure}[t]{0.1975\textwidth}
          \includegraphics[width=\linewidth, trim=5 0 4.5 0]{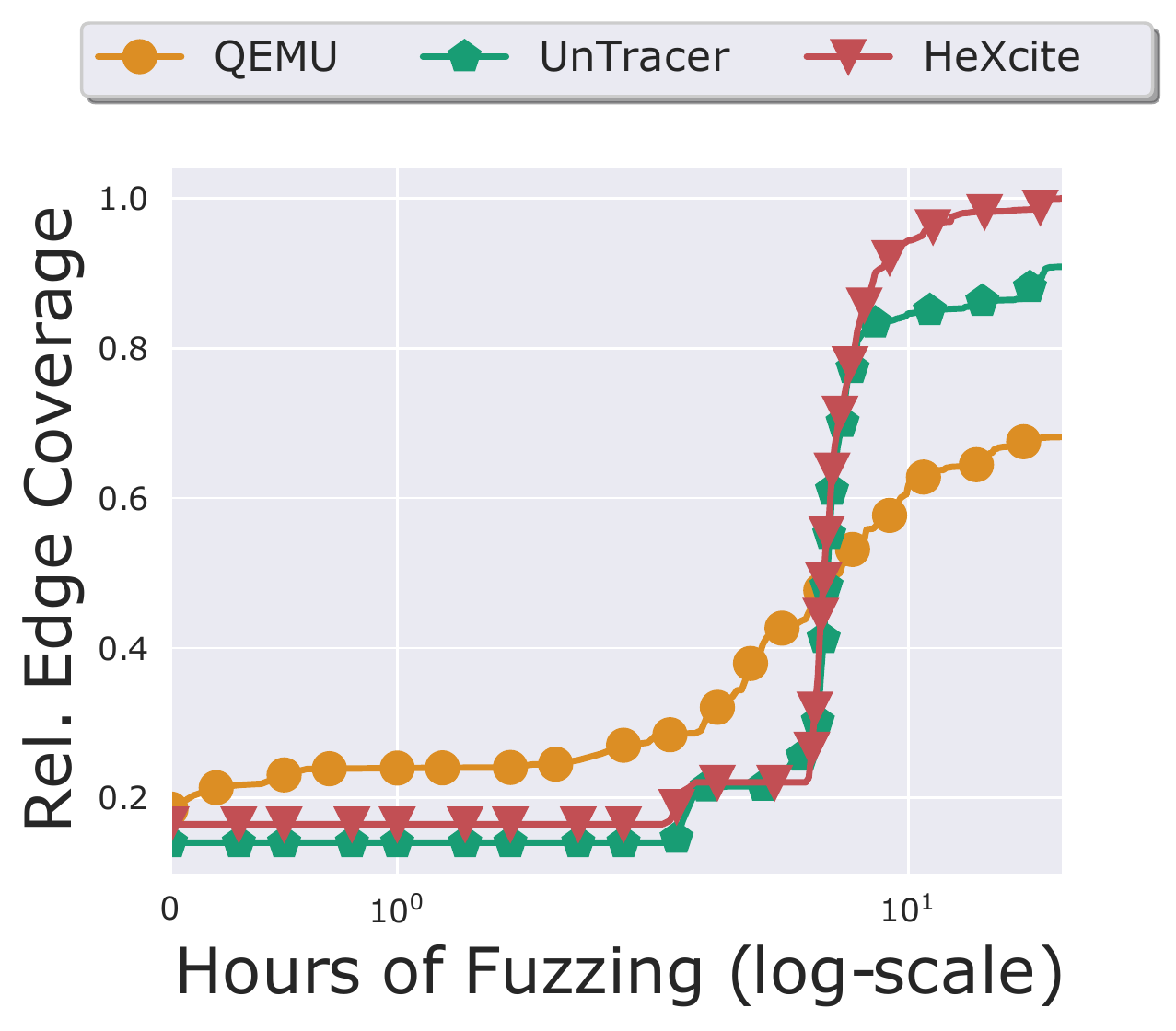}  
          \vspace{-0.4cm}
          \caption{\texttt{unrar}}
          \label{fig:coverage:edge:E}
     \end{subfigure}
    \vspace{-0.4cm}
    \caption{\textbf{\platname's mean code coverage over time} relative to all supported tracing approaches per benchmark. We log-scale the trial duration (24 hours) to more clearly show the end-of-fuzzing coverage divergence.}
    \label{fig:coverage:edge}
    \vspace{-0.1cm}
    \hrulefill
    \vspace{0.2cm}
\end{figure*}

\begin{table*}[t!]
\scriptsize
    \centering
    
    %
    \begin{tabular}[b]{l | c c | c c c c c c c c }
        \specialrule{.1em}{0em}{0em} 
        \hline
        
        \mystrut
        \multirow{3}{*}{\textbf{Binary}} & 
        \multicolumn{2}{c|}{{\notsosmall{vs. \emph{Coverage-guided Tracing}}}} &
        \multicolumn{8}{c}{{\notsosmall{vs. Binary- and Source-level \emph{Always-on Tracing}}}} \\

        \mystrut 
        &
        \multicolumn{2}{c|}{\cellcolor{white}\textbf{{\platname\phantom{}/ UnTracer}}} &
        \multicolumn{2}{c}{\cellcolor{white}\textbf{{\platname\phantom{}/ QEMU}}} &
        \multicolumn{2}{c}{\cellcolor{white}\textbf{{\platname\phantom{}/ Dyninst}}} &
        \multicolumn{2}{c}{\cellcolor{white}\textbf{{\platname\phantom{}/ \notsotiny{RetroWrite}}}} &
        \multicolumn{2}{c}{\cellcolor{white}\textbf{{\platname\phantom{}/ Clang}}} \\
        
        & 
        \notsotiny{{{Rel. Cov}}} & 
        \notsotiny{MWU } &
        \notsotiny{{{Rel. Cov}}} & 
        \notsotiny{MWU } &
        \notsotiny{{{Rel. Cov}}} & 
        \notsotiny{MWU } &
        \notsotiny{{{Rel. Cov}}} & 
        \notsotiny{MWU } &
        \notsotiny{{{Rel. Cov}}} &
        \notsotiny{MWU } \\

        \hline\hline
        
        \rowcolor{gray!10}\texttt{jasper} &
            1.04 & {0.403} & 
            {1.71} & \textbf{{$<$0.001}} & 
            {1.77} & \textbf{{$<$0.001}} & 
            \xmarktab & \xmarktab & 
            1.01 & {0.209} \\
            
        \texttt{mjs} & 
            {1.05} & \textbf{{0.002}} & 
            {1.07} & \textbf{{$<$0.001}} &  
            {1.09} & \textbf{{$<$0.001}} & 
            {1.04} & \textbf{{0.001}} &
            1.01 & {0.231} \\

        \rowcolor{gray!10}\texttt{nasm} & 
            {1.06} & \textbf{{$<$0.001}} &
            {1.15} & \textbf{{$<$0.001}} &
            {1.17} & \textbf{{$<$0.001}} &
            \xmarktab & \xmarktab &
            {1.03} & \textbf{{$<$0.001}} \\

        \texttt{sam2p} & 
            {1.03} & \textbf{{0.003}} &
            \xmarktab & \xmarktab &
            {1.12} & \textbf{{$<$0.001}} &
            \xmarktab & \xmarktab &
            1.02 & {0.292}  \\

        \rowcolor{gray!10}\texttt{sfconvert} & 
            {1.04} & \textbf{{$<$0.001}} &
            \xmarktab & \xmarktab &
            1.00 & {0.057} &
            1.00 & {0.492} &
            {0.99} & {0.031}  \\

        \texttt{tcpdump} & 
            {1.11} & \textbf{{$<$0.001}} &
            {1.41} & \textbf{{$<$0.001}} &
            {1.16} & \textbf{{$<$0.001}} &
            {1.13} & \textbf{{$<$0.001}} &
            {1.08} & \textbf{{0.002}} \\

        \rowcolor{gray!10}\texttt{unrtf} & 
            {1.18} & \textbf{{0.002}} & 
            1.02 & {0.168} & 
            {1.03} & \textbf{{0.041}} & 
            {1.06} & \textbf{{0.002}} & 
            1.00 & {0.440} \\

        \texttt{yara} & 
            1.03  &  {0.057} & 
            {1.08} &  \textbf{{0.028}} & 
            {1.12} &  \textbf{{0.034}} & 
            {1.09} & \textbf{{0.034}}  & 
            0.95 & {0.061}   \\

        \rowcolor{gray!10}\texttt{lzturbo} & 
          {1.01} & \textbf{{$<$0.001}} &
          {1.06} & \textbf{{$<$0.001}} &
          \xmarktab & \xmarktab &
          \xmarktab & \xmarktab &
          \xmarktab & \xmarktab \\

        \texttt{pngout} & 
          {1.08} & \textbf{{0.001}} &
          {1.33} & \textbf{{$<$0.001}} &
          \xmarktab & \xmarktab &
          \xmarktab & \xmarktab &
          \xmarktab & \xmarktab \\

        \rowcolor{gray!10}\texttt{rar} & 
          {1.02} & \textbf{{0.004}} & 
          {1.02} & \textbf{{0.026}} &
          \xmarktab & \xmarktab & 
          \xmarktab & \xmarktab & 
          \xmarktab & \xmarktab  \\

        \texttt{unrar} & 
          {1.10} & \textbf{{0.005}} &
          {1.47} & \textbf{{$<$0.001}} & 
          \xmarktab & \xmarktab &
          \xmarktab & \xmarktab & 
          \xmarktab & \xmarktab  \\

        \hline
        \multirow{1}{*}{{\textbf{Mean Increase}}} &
        {\textbf{+6.2\%}} & &  
        {\textbf{+23.1\%}} & &  
        {\textbf{+18.1\%}} & &  
        {\textbf{+6.3\%}} & &  
        {\textbf{+1.1\%}} \\
        
        
    \end{tabular}
    \vspace{-0.00cm}
    \caption{\textbf{\platname's mean code coverage} relative to {U}nTracer, {Q}EMU, {D}yninst, {R}etrowrite, and AFL-{C}lang. \emph{\xmarktab} = the competing tracer is incompatible with the respective benchmark and hence omitted. Statistically significant improvements for \platname (i.e., Mann-Whitney U test $p < 0.05$) are \textbf{bolded}.}
    \label{tab:coverage:edge}
\vspace{-0.4cm}
\hrulefill{}
\end{table*}

\par\textbf{Infrastructure:}
We carry out all evaluations on the Microsoft Azure cloud infrastructure.
Each fuzzing trial is issued its own isolated Ubuntu 16.04 x86-64 virtual machine. 
Following Klees et al.'s~\cite{klees_evaluating_2018} standard we run 16$\times$24-hour trials per benchmark for each of the coverage-tracing approaches listed in \autoref{tab:configs}, amounting to over 2.4 years' of total compute time across our entire evaluation.
All benchmarks are instrumented on an Ubuntu 16.04 x86-64 desktop with a 6-core 3.50GHz Intel Core i7-7800x CPU and 64GB memory.
We repurpose the same system for all data post-processing.


\subsection{Q1: Coverage Evaluation}
To understand the trade-offs of adapting CGT to finer-grained coverage metrics, we first evaluate \platname's code and loop coverage against the block-coverage-only Coverage-guided Tracer UnTracer; as well as conventional always-on coverage-tracing approaches QEMU, Dyninst, RetroWrite, and AFL-Clang.
We detail our experimental setup and results below.

\subsubsection{\textbf{Code Coverage}}
\label{sec:eval:covg:edge}
We compare the code coverage of all tracing approaches in \autoref{tab:configs}.
We utilize AFL++'s Link Time Optimization (LTO) instrumentation~\cite{fioraldi_afl_2020} to build \emph{collision-free} edge-tracking versions of each binary; 
the same technique is applied to our four closed-source benchmarks (\autoref{tab:benchmarks}) with the help of the industry-standard binary-to-LLVM lifting tool McSema~\cite{dinaburg_mcsema_2014}.
We measure each trial's code coverage by replaying its test cases on the LTO binary using AFL's \texttt{afl-showmap}~\cite{zalewski_american_2017} utility and compute the average across all 16 trials.
\autoref{tab:coverage:edge} reports the average across all benchmark--tracer pairs as well as Mann-Whitney U significance scores at the $p=0.05$ significance level; and \autoref{fig:coverage:edge} shows the relative edge coverage over 24-hours for several benchmarks.

\par\textbf{Versus UnTracer:}
As \autoref{tab:coverage:edge} shows, \platname surpasses UnTracer in total coverage across all benchmarks by \textbf{1--18\%} for a mean improvement of \textbf{6.2\%}, with statistically higher coverage on 10 of 12 benchmarks.
The impact of coverage granularity on CGT is significant; besides seeing the worst coverage on \texttt{unrtf} (\autoref{fig:coverage:edge:C}) and \texttt{sfconvert}, block-only coverage UnTracer is bested by AFL-Clang on all 8 open-source benchmarks, demonstrating that sheer speed is not enough to overcome a sacrifice in code coverage---whereas \platname's \emph{coverage-preserving} CGT averages the highest overall code coverage in our entire evaluation.

\par\textbf{Versus binary-only always-on tracing:}
We see that \platname achieves a mean \textbf{23.1\%}, \textbf{18.1\%}, and \textbf{6.3\%} higher code coverage over binary-only always-on tracers QSYM, Dyninst, and RetroWrite (respectively), with statistically significant improvements on all but one binary per comparison (\texttt{yara} for QEMU, and \texttt{sfconvert} for Dyninst and RetroWrite).
For \texttt{sfconvert} in particular, we find that all tracers' runs are dominated by timeout-inducing inputs, causing each to see roughly equal execution speeds, and hence, code coverage.
While we expect that timeout-laden binaries are less likely to see benefit from CGT in general, overall, \platname's balance of fine-grained coverage \emph{and} speed easily rank it the highest-coverage binary-only tracer.

\par\textbf{Versus source-level always-on tracing:}
Across all eight open-source benchmarks \platname averages \textbf{1.1\%} higher coverage than AFL's source-level tracing, AFL-Clang.
Despite having statistically worse coverage on \texttt{sfconvert} (due to its heavy timeouts), \platname's coverage is statistically better or identical to AFL-Clang's on 7/8 benchmarks, confirming that coverage-preserving CGT brings coverage tracing \emph{at~least~as~effective~as} source-level tracing---to binary-only fuzzing use cases.

\begin{table}[!h]
\footnotesize    
    \centering
    
    \begin{tabular}[b]{c | c | c}
        \specialrule{.1em}{0em}{0em} 
        \hline
        \mystrut
        \multirow{3}{*}{\textbf{Binary}} & 
            \multirow{2}{*}{\textbf{\scriptsize{\shortstack{\platname \\ / UnTracer}}}} &
            \multirow{2}{*}{\textbf{\scriptsize{\shortstack{\platname \\ / Clang}}}} \\
            
        & & \\
        & \notsotiny{Rel. LoopCov} & \notsotiny{Rel. LoopCov} \\
            
        \hline\hline
        
        \rowcolor{gray!10}\texttt{jasper} & 
            1.56 & 1.14 \\
            
        \texttt{mjs} & 
            3.61 & 1.06 \\

        \rowcolor{gray!10}\texttt{nasm} & 
            2.54 & 1.85 \\

        \texttt{sam2p} & 
            1.05 & 1.19  \\

        \rowcolor{gray!10}\texttt{sfconvert} & 
            1.89 & 2.56  \\

        \texttt{tcpdump} & 
            1.21 & 1.39  \\

        \rowcolor{gray!10}\texttt{unrtf} & 
            3.54 & 0.73  \\

        \texttt{yara} & 
            2.98 & 0.95  \\

        \hline
        \multirow{1}{*}{\scriptsize{\textbf{Mean Increase}}} &
            \textbf{+130\%} & \textbf{+36\%} \\
        
    \end{tabular}
    \vspace{-0.00cm}
    \caption{\textbf{\platname's mean loop coverage} (i.e., average maximum consecutive iterations capped at 128) relative to block-only CGT UnTracer and the source-level conventional tracer AFL-Clang.}
    \label{tab:coverage:loop}
\vspace{-0.4cm}
\hrulefill{}
\end{table}
\begin{figure}[h!]
    \centering
    \tiny
     \begin{subfigure}[t]{0.237\textwidth}
          \includegraphics[width=\linewidth]{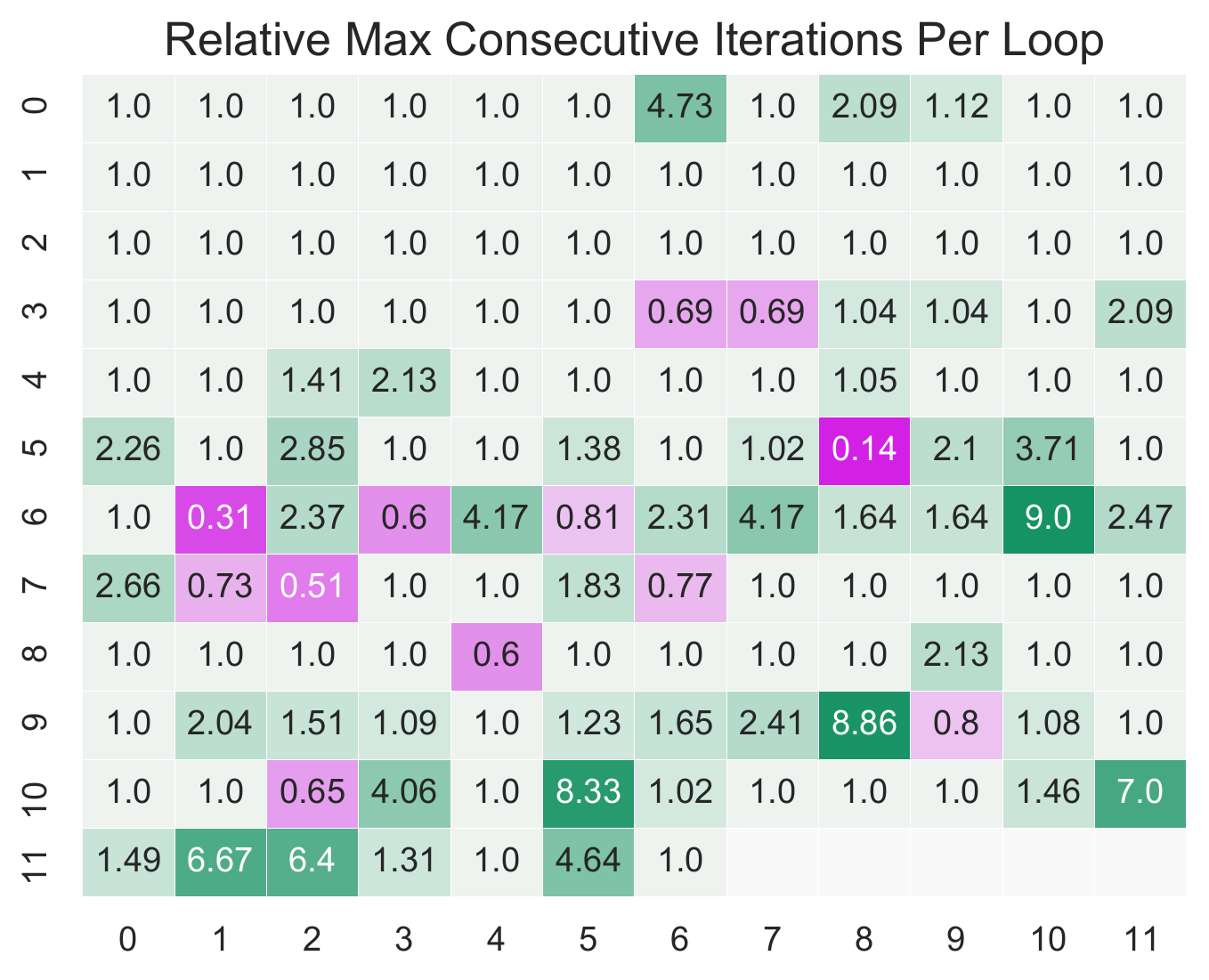}
          \vspace{-0.4cm}
          \caption{ \texttt{jasper}}
          \label{fig:coverage:loop:A}
     \end{subfigure}
     \begin{subfigure}[t]{0.237\textwidth}
          \includegraphics[width=\linewidth]{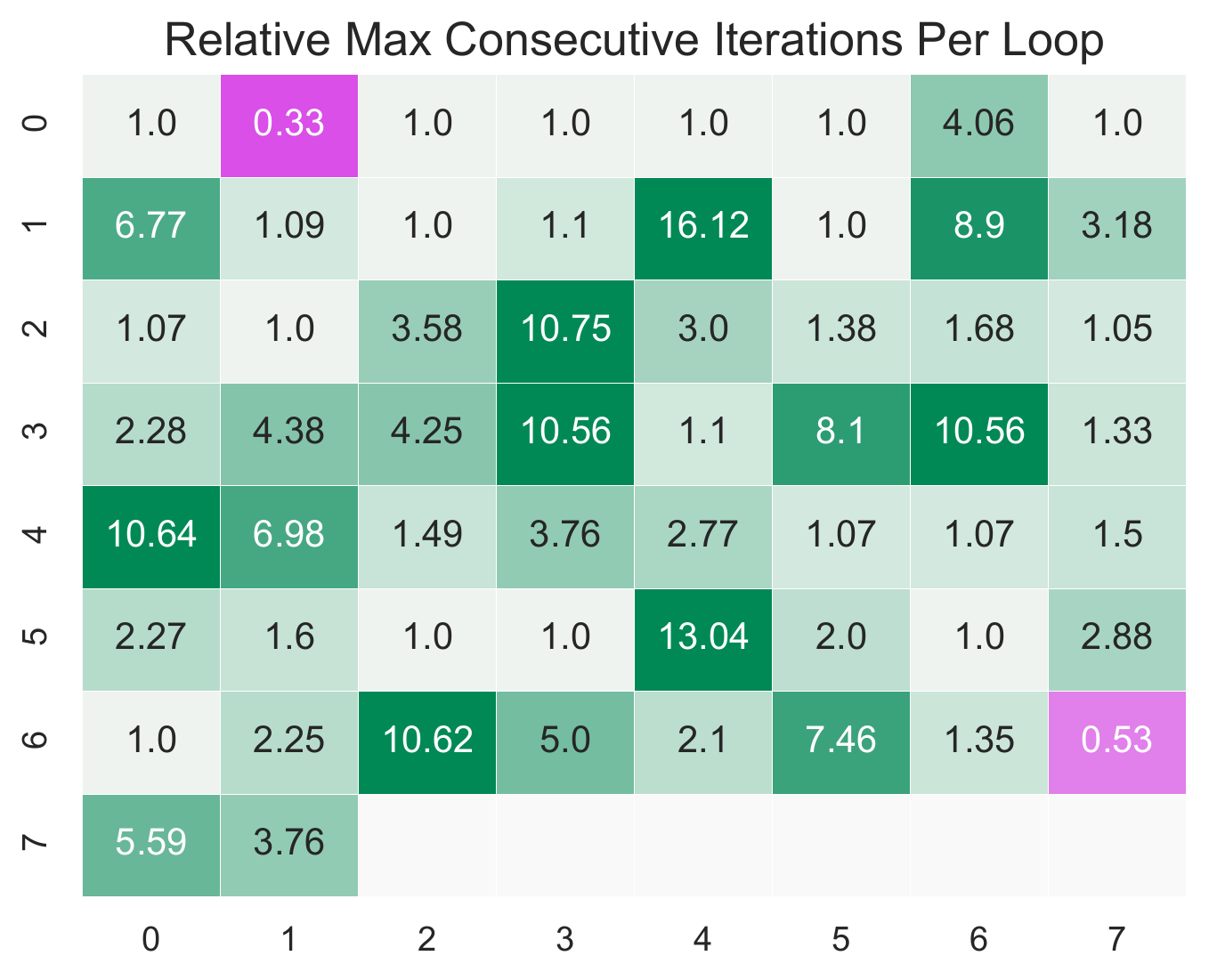}  
          \vspace{-0.4cm}
          \caption{\texttt{mjsbin}}
          \label{fig:coverage:loop:B}
     \end{subfigure}          
     \begin{subfigure}[t]{0.237\textwidth}
          \includegraphics[width=\linewidth]{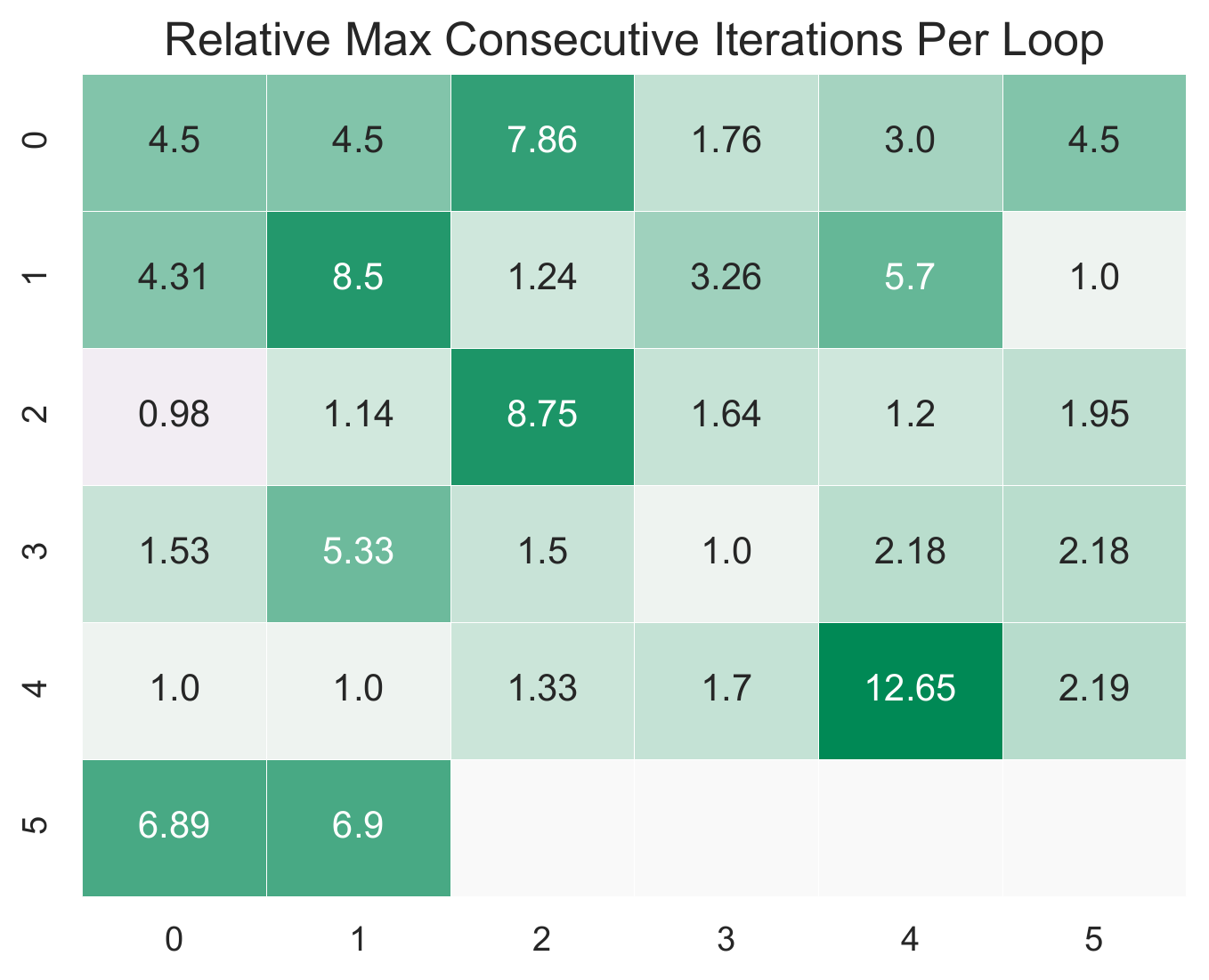}  
          \vspace{-0.4cm}
          \caption{\texttt{unrtf}}
          \label{fig:coverage:loop:C}          
     \end{subfigure}
     \begin{subfigure}[t]{0.237\textwidth}
          \includegraphics[width=\linewidth]{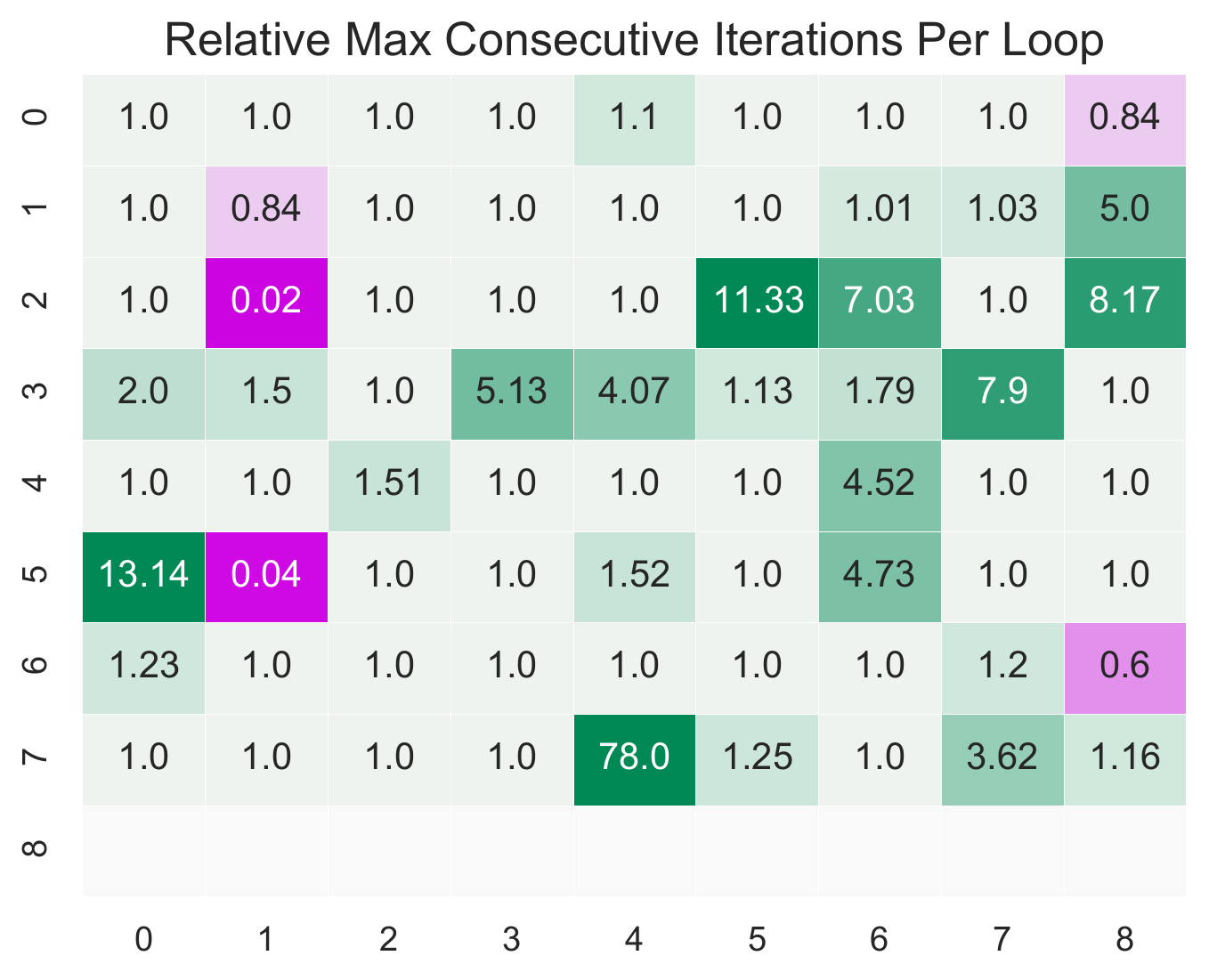}
          \vspace{-0.4cm}
          \caption{ \texttt{yara}}
          \label{fig:coverage:loop:D}
     \end{subfigure}

    \vspace{-0.4cm}
    \caption{\textbf{\platname's mean loop coverage} relative to UnTracer. Each box represents a mutually-covered loop, with values indicating the mean maximum consecutive iterations (capped at 128 total iterations to match AFL) over all 16 trials. \emph{Green} and \emph{pink} shading indicate a higher relative loop coverage for \platname and UnTracer (respectively), while \emph{grey} indicates no change. }
    \label{fig:coverage:loop}
    \vspace{-0.1cm}
    \hrulefill
\end{figure}

\begin{figure*}[!t]
        \centering
          \includegraphics[width=\linewidth]{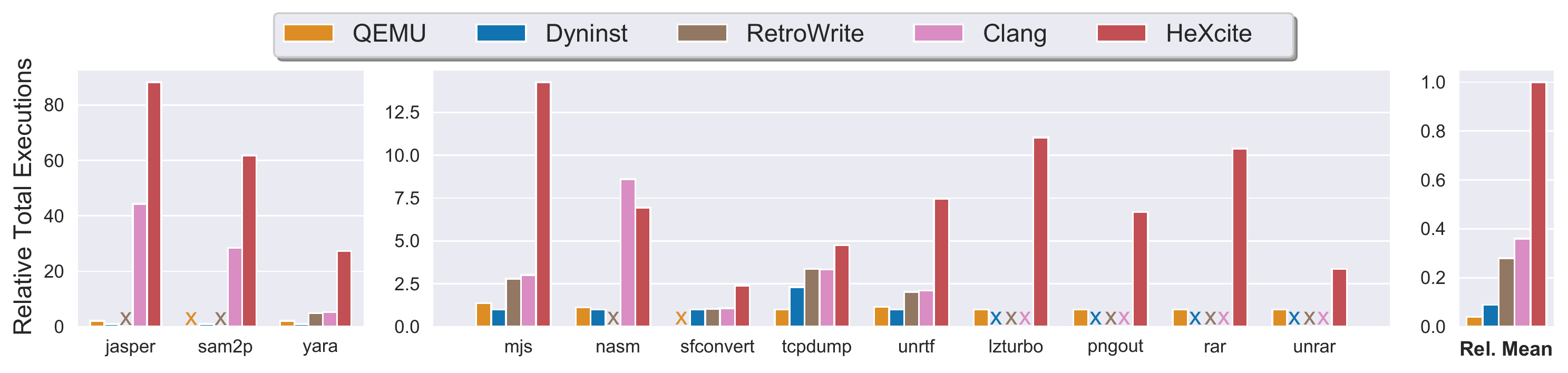}  
        \vspace{-0.9cm}
        \caption{\textbf{\platname's mean throughput relative to conventional coverage tracers}. We normalize throughput to the worst-performing tracer per benchmark, and compute each tracer's mean performance relative to \platname's across all benchmarks (shown in the rightmost plot).
        For each benchmark we omit incompatible tracers (denoted by a colored \textbf{\xmarktab}).
        \emph{All comparisons to \platname yield a statistically significant difference} (i.e., Mann-Whitney U test $p$ < 0.05).}
        \label{fig:perf}
        \vspace{-0.275cm}
        \hrulefill
        \vspace{0.2cm}
\end{figure*}

\subsubsection{\textbf{Loop Coverage}}
\label{sec:eval:covg:loop}
To determine if coverage-preserving CGT is more effective at covering code loops, we develop a custom LLVM instrumentation pass to report the maximum consecutive iterations per loop per trial.
Despite our success in lifting our closed-source benchmarks to add edge-tracking instrumentation (\autoref{sec:eval:covg:edge}), none of our binary-to-LLVM lifters (McSema, rev.ng, RetDec, reopt, llvm-mctoll, or Ghidra-to-LLVM) succeeded in recovering the loop metadata necessary for our LLVM loop transformation to work; thus our loop analysis is restricted to our eight open-source benchmarks. 

We compare \platname to UnTracer and AFL-Clang as they support all eight open-source benchmarks (and hence omit QEMU, Dyninst and RetroWrite which only support a few).
We compute each loop's mean from the maximum consecutive iterations for all trials per benchmark--tracer pair, capping iterations at 128 as AFL omits hit counts beyond this range.
\autoref{tab:coverage:loop} reports \platname's mean coverage across all loops for each binary relative to UnTracer and AFL-Clang;
and \autoref{fig:coverage:loop} shows a heatmap of \platname's per-loop coverage relative to UnTracer's for several benchmarks.

\par\textbf{Versus UnTracer:}
As~\autoref{tab:coverage:loop} shows, \platname's bucketed unrolling brings \textbf{130\%} higher loop penetration coverage over UnTracer.
We see that UnTracer beats \platname on a minutia of loops per benchmark (\autoref{fig:coverage:loop})---expectedly---as its inability to track loop progress inevitably constrains fuzzing to exploring the same few loops trial after trial.
We find that \platname queues over \textbf{2$\times$} as many test cases, thus showing that its loop-progress-aware coverage leads fuzzing to sacrifice focusing on the same few loops in favor of a \emph{higher~diversity} of loops per binary.

\par\textbf{Versus source-level always-on tracing:}
We see that, on average, \platname attains a \textbf{36\%} higher loop coverage than source-level always-on tracing with AFL-Clang.
Though this improvement is modest, these results show that bucketed unrolling outperforms conventional coverage tracing's exhaustive (i.e., on every basic block) hit count tracking---yet only instruments loop headers. 
While we posit that bucketed unrolling has further optimization potential (e.g., halving the number of buckets, selective insertion, etc.), we leave exploring this trade-off space to future work.

\vspace{0.1cm}
\stevebox{}{
\small
\textbf{Q1:} 
Jump mistargeting and bucketed unrolling enable Coverage-preserving CGT to achieve the highest overall coverage versus \emph{block-only} CGT---as well as conventional binary \emph{and} source-level tracing.
}

\subsection{Q2: Performance  Evaluation}
\label{sec:eval:perf}
To measure the impacts of finer-grained coverage on CGT performance, we perform a piece-wise evaluation of the fuzzing test case throughput (i.e., mean total test cases processed in 24-hours) of \platname's \emph{edge} (via jump mistargeting) and \emph{full} (jump mistargeting + bucketed unrolling) coverage versus UnTracer's block-only coverage, shown in \autoref{tab:perf:piecewise}. 
To ascertain where coverage-preserving CGT's performance stands with respect to always-on tracing, we further evaluate \platname's best-case throughput alongside the leading binary- and source-level coverage tracers QEMU, Dyninst, RetroWrite, and AFL-Clang, shown in \autoref{fig:perf}.

\begin{table}[!h]
\footnotesize   
    \centering
    
    \begin{tabular}[b]{l | M{0.6cm} M{0.63cm} M{0.6cm} M{0.63cm} M{0.6cm} M{0.63cm} }
        \specialrule{.1em}{0em}{0em} 
        \hline
        \mystrut
        \multirow{2}{*}{\textbf{Binary}} & 
            \multicolumn{2}{c}{\textbf{{Edge / Block}}} &
            \multicolumn{2}{c}{\textbf{{Full / Block}}} &
            \multicolumn{2}{c}{\textbf{{Best / Block}}} \\
        
        & \notsotiny{Rel.\phantom{.}Perf} & \notsotiny{MWU} & 
        \notsotiny{Rel.\phantom{.}Perf} & \notsotiny{MWU} & 
        \notsotiny{Rel.\phantom{.}Perf} & \notsotiny{MWU} \\
            
            
        \hline\hline
        
        \rowcolor{gray!10}\texttt{jasper} & 
            {0.52} & 
            {$<$0.001} & 
            {0.54} & 
            {$<$0.001} & 
            \cellcolor{gray!30}{0.54} & 
            \cellcolor{gray!30}{$<$0.001} \\
            
        \texttt{mjs} & 
            {0.93} & 
            {0.046} & 
            {0.65}  & 
            {$<$0.001} & 
            \cellcolor{gray!30}{0.93} & 
            \cellcolor{gray!30}{0.046} \\

        \rowcolor{gray!10}\texttt{nasm} &
            {1.46} & 
            {$<$0.001} &
            {2.61} & 
            {$<$0.001} &
            \cellcolor{gray!30}{2.61} & 
            \cellcolor{gray!30}{$<$0.001} \\

        \texttt{sam2p} & 
            0.99 & 
            {0.433} & 
            1.07 & 
            {0.090} & 
            \cellcolor{gray!30}1.07 & 
            \cellcolor{gray!30}{0.090} \\

        \rowcolor{gray!10}\texttt{sfconvert} & 
            {1.06} & 
            {$<$0.001} & 
            {1.24} & 
            {$<$0.001} & 
            \cellcolor{gray!30}{1.24} & 
            \cellcolor{gray!30}{$<$0.001} \\

        \texttt{tcpdump} & 
            0.96 & 
            {0.150} & 
            {0.64} & 
            {$<$0.001} & 
            \cellcolor{gray!30}0.96 & 
            \cellcolor{gray!30}{0.150} \\

        \rowcolor{gray!10}\texttt{unrtf} & 
            1.04 & 
            {0.332} & 
            {0.78} & 
            {$<$0.001} & 
            \cellcolor{gray!30}1.04 & 
            \cellcolor{gray!30}{0.332} \\

        \texttt{yara} & 
            0.97 & 
            {0.125} & 
            {0.18} & 
            {$<$0.001} & 
            \cellcolor{gray!30}0.97 & 
            \cellcolor{gray!30}{0.125} \\

        
        \rowcolor{gray!10}\texttt{lzturbo} & 
            0.74 & 
            {0.292} & 
            0.82 & 
            {0.448} & 
            \cellcolor{gray!30}0.82 & 
            \cellcolor{gray!30}{0.448} \\

        \texttt{pngout} & 
            {1.02} & 
            {0.002} & 
            0.99 & 
            {0.332} & 
            \cellcolor{gray!30}{1.02} & 
            \cellcolor{gray!30}{0.002} \\		
            
        \rowcolor{gray!10}\texttt{rar} & 
            1.01 & 
            {0.492} & 
            {0.68} & 
            {$<$0.001} & 
            \cellcolor{gray!30}1.01 & 
            \cellcolor{gray!30}{0.492} \\

        \texttt{unrar} & 
            0.97 & 
            {0.188} & 
            {0.90} & 
            {0.047} & 
            \cellcolor{gray!30}0.97 & \cellcolor{gray!30}{0.188} \\
		        
        \hline
        \multirow{1}{*}{\scriptsize{\textbf{Mean Rel. Perf.}}} &
            \textbf{97\%} & & \textbf{92\%} & & \textbf{110\%} & \\
            
    \end{tabular}
    \vspace{-0.00cm}
    \caption{\textbf{Performance trade-offs of different CGT coverage granularities}. We compute mean throughputs for three \platname coverage granularities (edge, full, and the best of both) relative to UnTracer's \emph{block}-only granularity. }
    \label{tab:perf:piecewise}
\vspace{-0.4cm}
\hrulefill{}
\end{table}

\begin{figure*}[!t]
    \centering
    \tiny
    \begin{subfigure}[t]{0.235\textwidth}
          \includegraphics[trim=0 0 0 0, width=\linewidth]{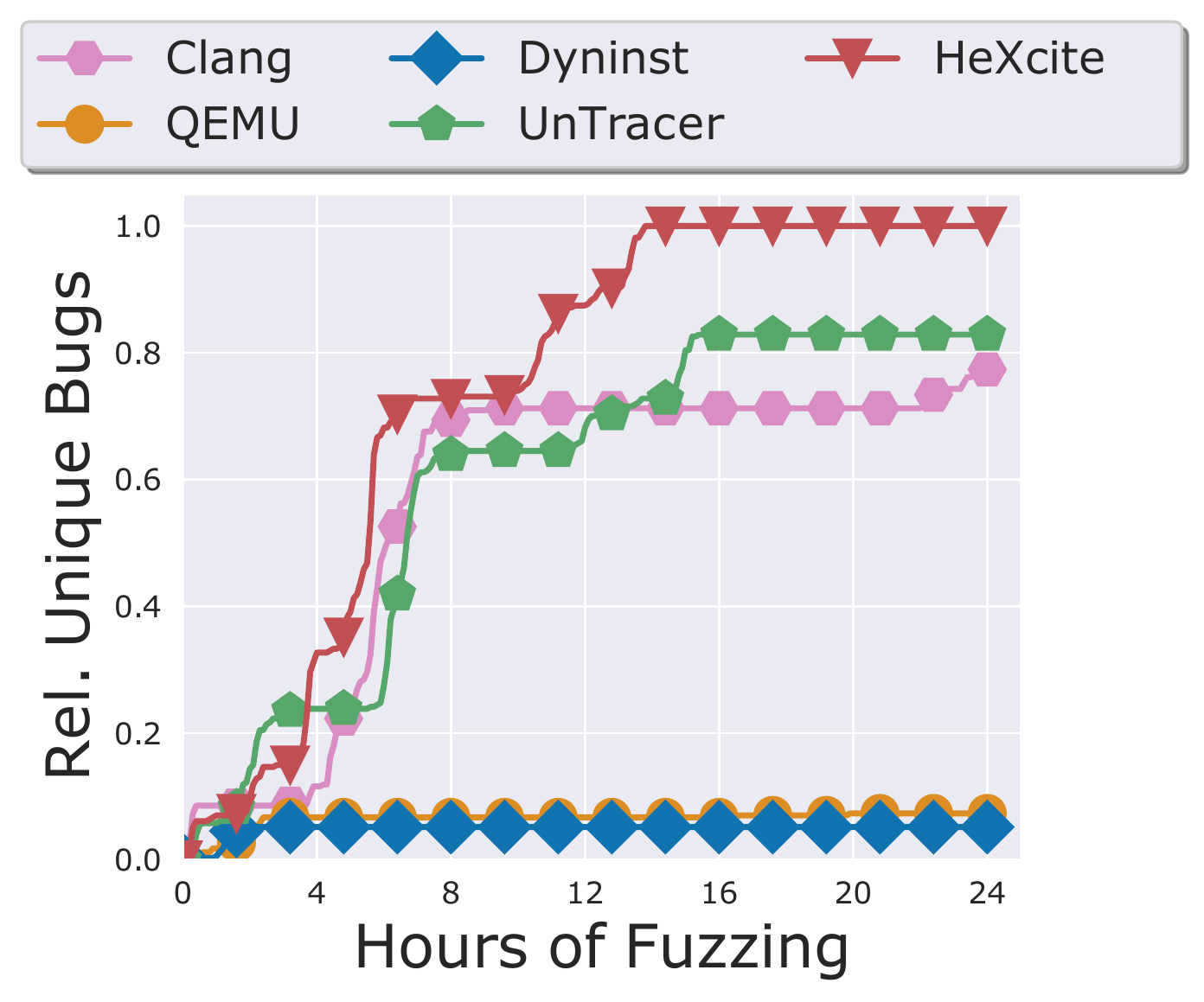}
          \vspace{-0.4cm}
          \caption{ \texttt{nasm}}
          \label{fig:rw:crash:A}
     \end{subfigure}
     \begin{subfigure}[t]{0.255\textwidth}
          \includegraphics[trim=0 0 0 0, width=\linewidth]{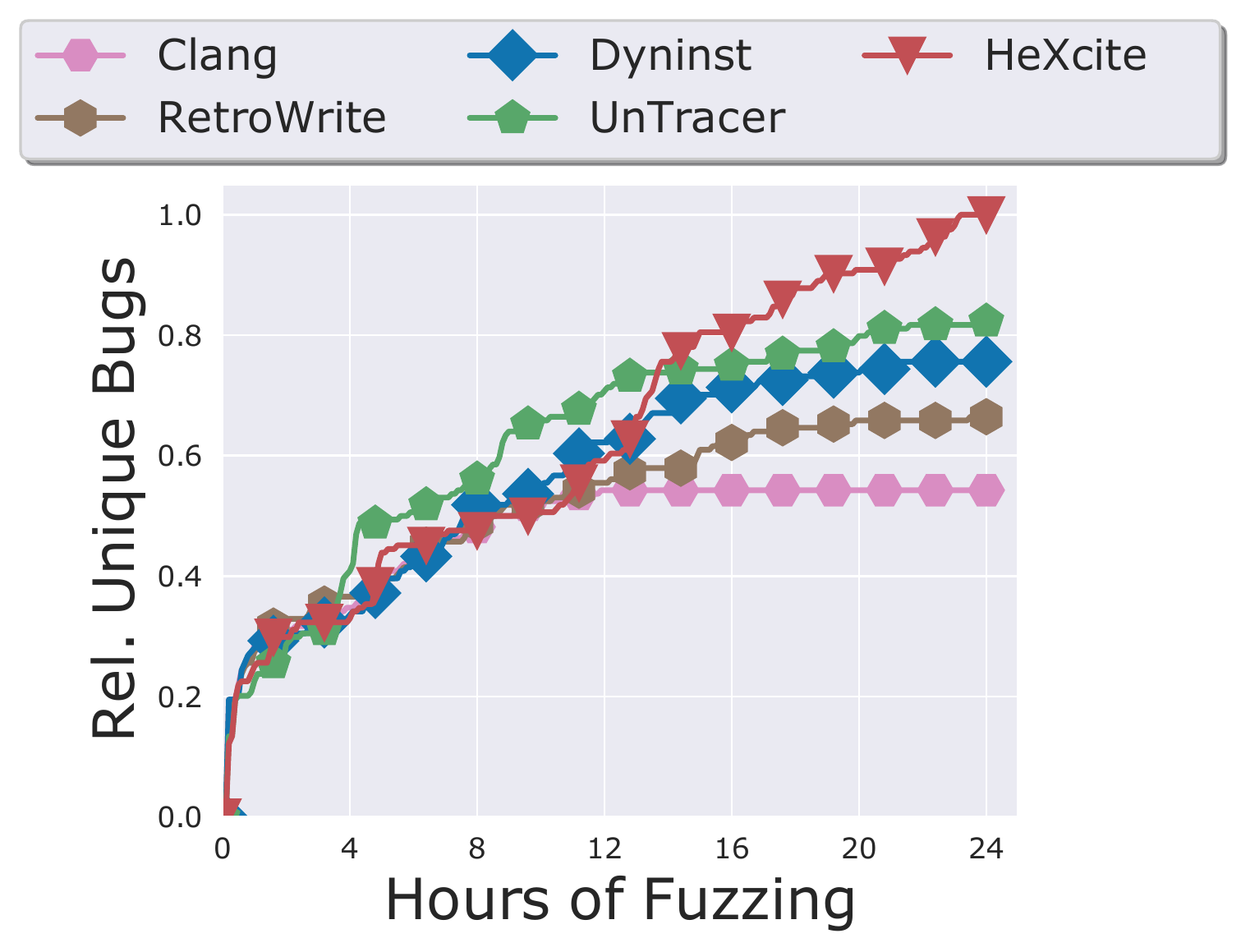}
          \vspace{-0.4cm}
          \caption{\texttt{sfconvert}}
          \label{fig:rw:crash:B}
     \end{subfigure}
     \begin{subfigure}[t]{0.25\textwidth}
          \includegraphics[trim=0 0 0 0, width=\linewidth]{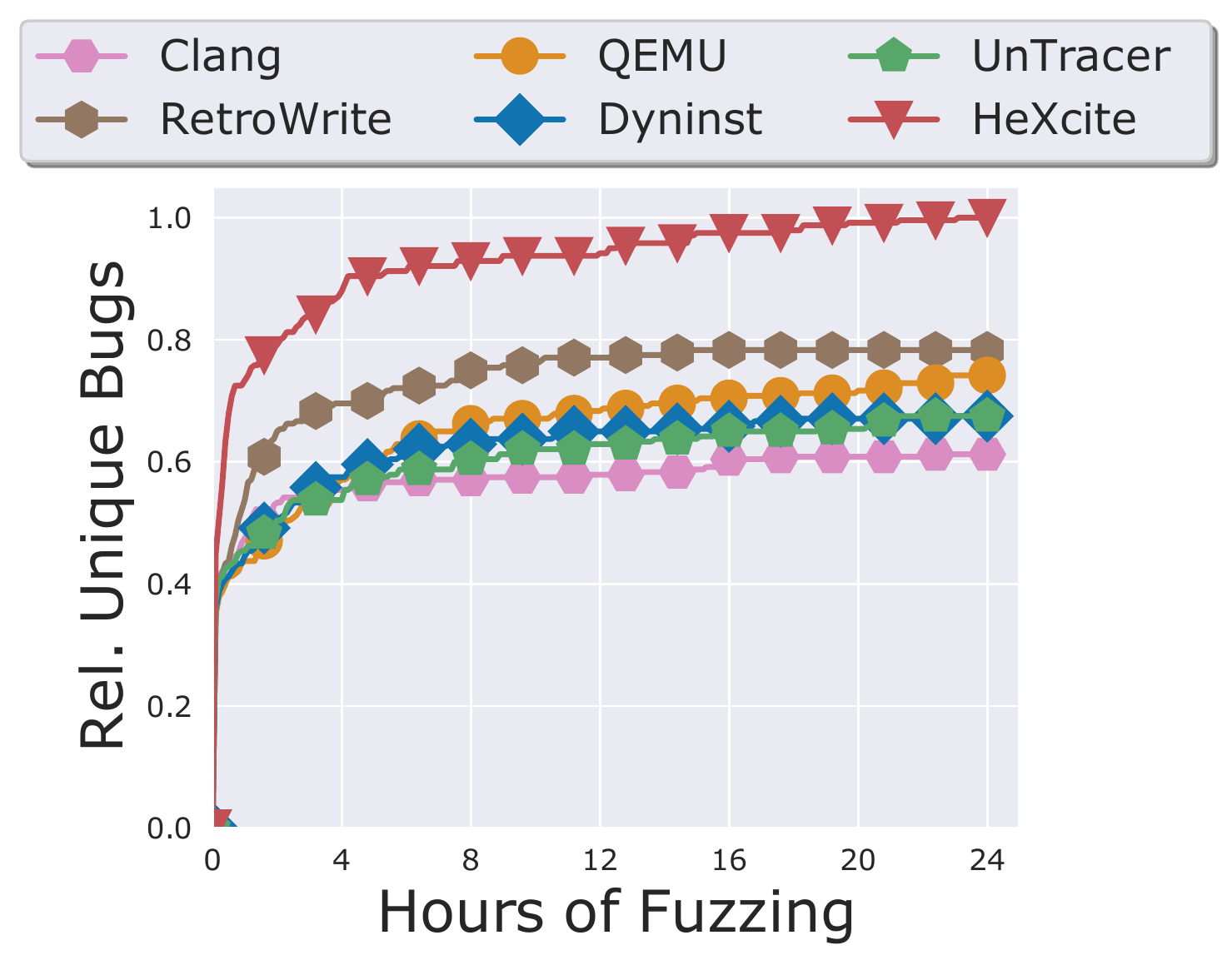}  
          \vspace{-0.4cm}
          \caption{\texttt{unrtf}}
          \label{fig:rw:crash:C}
     \end{subfigure}
     \begin{subfigure}[t]{0.24\textwidth}
          \includegraphics[trim=0 0 0 0, width=\linewidth]{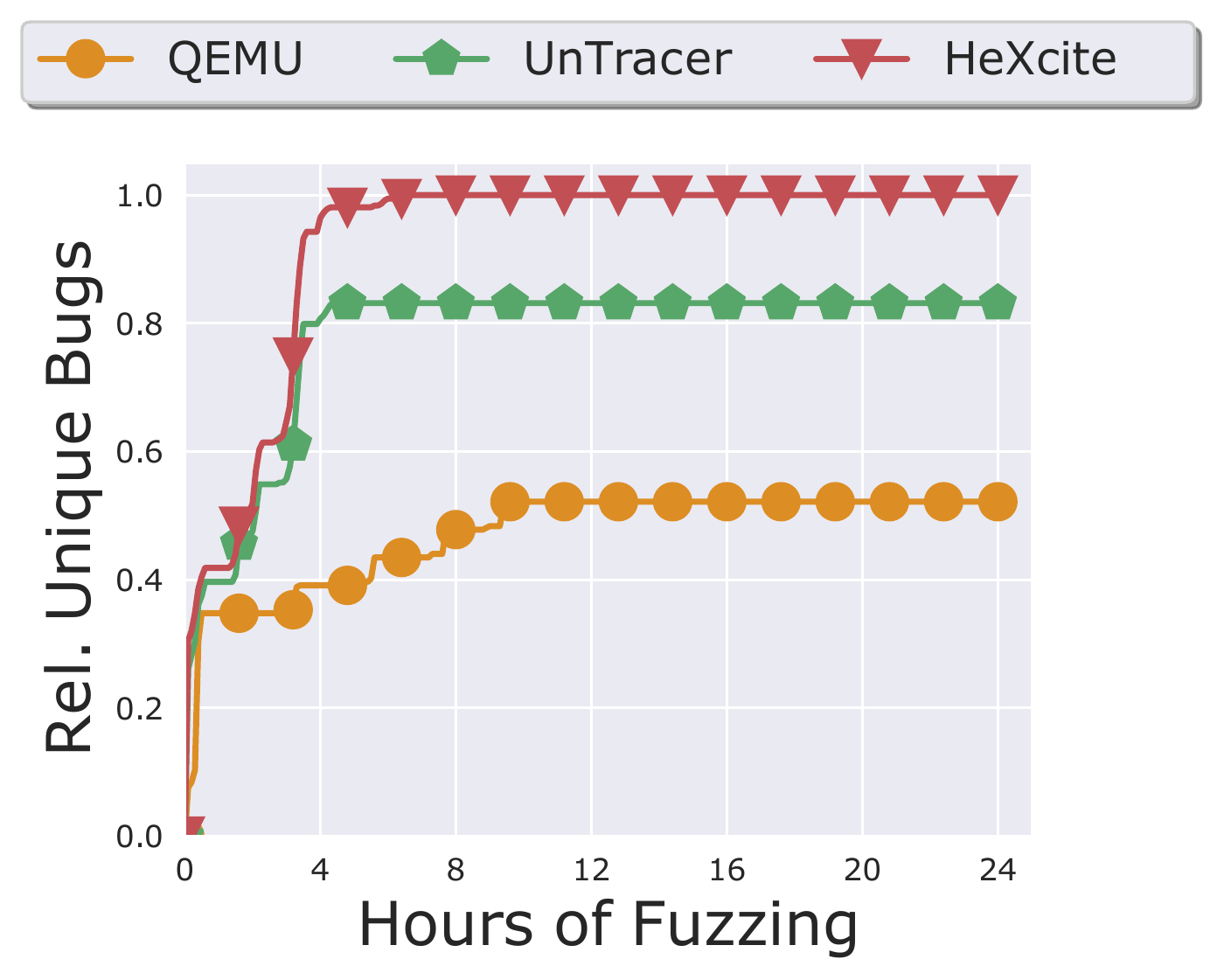}  
          \vspace{-0.4cm}
          \caption{\texttt{pngout}}
          \label{fig:rw:crash:D}
     \end{subfigure}
    \vspace{-0.4cm}
    \caption{\textbf{\platname's mean unique bugs over time} relative to all supported tracing approaches per benchmark.}    
    
    \label{fig:bugs}
    \vspace{-0.1cm}
    \hrulefill
    \vspace{0.2cm}
\end{figure*}
\begin{table*}[!h]
\scriptsize   
    \centering
    
    \begin{tabular}[b]{p{1.55cm} | M{0.6cm}M{0.6cm}M{0.8cm} | M{0.6cm}M{0.6cm}M{0.7cm} M{0.7cm}M{0.6cm}M{0.5cm} M{0.6cm}M{0.6cm}M{0.7cm} M{0.6cm}M{0.6cm}M{0.6cm}}
        \specialrule{.1em}{0em}{0em} 
        \hline
        \mystrut
        \multirow{4}{*}{\textbf{Binary}} & 
        \multicolumn{3}{c|}{{\notsosmall{vs. \emph{Coverage-guided Tracing}}}} &
        \multicolumn{12}{c}{{\notsosmall{vs. Binary- and Source-level \emph{Always-on Tracing}}}} \\

        \mystrut 
        &
        \multicolumn{3}{c|}{\cellcolor{white}\textbf{{\platname\phantom{}/ UnTracer}}} &
        \multicolumn{3}{c}{\cellcolor{white}\textbf{{\platname\phantom{}/ QEMU}}} &
        \multicolumn{3}{c}{\cellcolor{white}\textbf{{\platname\phantom{}/ Dyninst}}} &
        \multicolumn{3}{c}{\cellcolor{white}\textbf{{{\platname\phantom{}/ \notsotiny{RetroWrite}}}}} &
        \multicolumn{3}{c}{\cellcolor{white}\textbf{{\platname\phantom{}/ Clang}}} \\
        
        & 
        \notsotiny{{{Rel. Crash}}} & 
        \notsotiny{{{Rel. Bugs}}} & 
        \notsotiny{MWU } &
        \notsotiny{{{Rel. Crash}}} & 
        \notsotiny{{{Rel. Bugs}}} & 
        \notsotiny{MWU } &
        \notsotiny{{{Rel. Crash}}} & 
        \notsotiny{{{Rel. Bugs}}} & 
        \notsotiny{MWU } &
        \notsotiny{{{Rel. Crash}}} & 
        \notsotiny{{{Rel. Bugs}}} & 
        \notsotiny{MWU } &
        \notsotiny{{{Rel. Crash}}} & 
        \notsotiny{{{Rel. Bugs}}} & 
        \notsotiny{MWU } \\
 
        \hline
        \hline
        \rowcolor{gray!10}\texttt{jasper} & 
        {1.40} & 0.97 & {0.241} &
        {25.32} &  {19.92} & \textbf{{$<$0.001}} &
        {42.50} &  {37.00} & \textbf{{$<$0.001}} &
        \xmarktab &  \xmarktab & \xmarktab & 
        {1.31} &  1.12 & {0.216}\\
                
        \texttt{mjs} & 
        {1.37} & 1.02 & {0.462} &
        {17.33} & {6.71} &  \textbf{{$<$0.001}} &
        {12.27} & {3.38} &  \textbf{{$<$0.001}} &
        {5.92} & {1.84} &  \textbf{{$<$0.001}} &
        {5.22} & {1.82} & \textbf{{$<$0.001}} \\
                
        \rowcolor{gray!10}\texttt{nasm} & 
        {1.99} & {1.21} & \textbf{{$<$0.001}} &
        {20.03} & {13.63} & \textbf{{$<$0.001}} &
        {18.93} & {19.24} & \textbf{{$<$0.001}} &
        \xmarktab & \xmarktab & \xmarktab &
        {1.74} & {1.27} &  \textbf{{$<$0.001}}\\
                
        \texttt{sam2p} & 
        {1.43} & 1.05 & {0.447} &
        \xmarktab & \xmarktab &  \xmarktab &
        {2.24} & {1.36} & \textbf{{$<$0.001}} &
        \xmarktab & \xmarktab & \xmarktab &
        {1.32} & {1.21} & \textbf{{0.018}} \\
                
        \rowcolor{gray!10}\texttt{sfconvert} & 
        {1.52} & {1.23} & \textbf{{$<$0.001}} &
        \xmarktab & \xmarktab &  \xmarktab &
        {1.42} & {1.35} & \textbf{{$<$0.001}} &
        {1.56} & {1.53} & \textbf{{$<$0.001}} &
        {1.78} & {1.88} & \textbf{{$<$0.001}} \\
                
        \texttt{tcpdump} & 
        {1.28} & 1.04 & {0.212} &
        {2.43} & {1.64} & \textbf{{$<$0.001}} &
        {1.91} & {1.27} & \textbf{{$<$0.001}} &
        {1.29} & {1.09} & \textbf{{0.048}} &
        1.01 & 1.05 & {0.084}\\
        
        \rowcolor{gray!10}\texttt{unrtf} & 
        {1.88} & {1.48} & \textbf{{$<$0.001}} &
        {1.37} & {1.35} & \textbf{{0.001}} &
        {1.67} & {1.46} & \textbf{{$<$0.001}} &
        {1.18} & {1.28} & \textbf{{0.001}} &
        {1.10} & {1.63} & \textbf{{$<$0.001}} \\
                
        \texttt{yara} & 
        0.72 & 1.02 &  {0.215} &
        {16.80} &  {2.34} & \textbf{{$<$0.001}} &
        {22.49} &  {2.89} & \textbf{{$<$0.001}} &
        {12.58} &  {2.05} & \textbf{{$<$0.001}} &
        {10.47} &  {1.72} & \textbf{{$<$0.001}} \\
                
        
        \rowcolor{gray!10}\texttt{pngout} & 
        {1.27} & {1.36} & \textbf{{$<$0.001}} &
        {2.49} & {2.17} & \textbf{{$<$0.001}} &
        \xmarktab & \xmarktab &  \xmarktab &
        \xmarktab & \xmarktab &  \xmarktab &
        \xmarktab & \xmarktab & \xmarktab \\
                
        \texttt{unrar} & 
        1.25 & 0.80 &  {0.279} &
        {2.00} & {2.00} & \textbf{{0.039}} &
        \xmarktab & \xmarktab &  \xmarktab &
        \xmarktab & \xmarktab &  \xmarktab &
        \xmarktab & \xmarktab &  \xmarktab\\
                
        \hline
                
        \multicolumn{1}{l|}{{\textbf{Mean Increase}}} & 
        \textbf{+41\%} & \textbf{+12\%} & &  
        \textbf{+997\%} & \textbf{+521\%} & &   
        \textbf{+1193\%} & \textbf{+749\%} & &   
        \textbf{+350\%} & \textbf{+56\%} & &  
        \textbf{+199\%} & \textbf{+46\%} & \\ 
                

    \end{tabular}
    \vspace{-0.0cm}
    \caption{\textbf{\platname's mean crashes and bugs} relative to UnTracer, QEMU, Dyninst, RetroWrite, and AFL-Clang. We omit \texttt{lzturbo} and \texttt{rar} as none trigger any crashes for them. \emph{\xmarktab} = the tracer is incompatible with the respective benchmark and hence omitted. Statistically significant improvements in \emph{mean bugs found} for \platname (i.e., Mann-Whitney U test $p < 0.05$) are \textbf{bolded}.}
    \label{tab:bugs}
\vspace{-0.4cm}
\hrulefill{}
\end{table*}

\par\textbf{Versus UnTracer:}
As \autoref{tab:perf:piecewise} shows, incorporating edge coverage in CGT incurs a mean throughput slowdown of \textbf{3\%}, while supporting full coverage (i.e., edges \emph{and} counts) sees a slightly higher slowdown of \textbf{8\%}.
However, as the experiments in \autoref{sec:eval:covg:edge} and \autoref{sec:eval:covg:loop} show, coverage-preserving CGT attains the highest edge and loop coverage of all tracers in our evaluation---offsetting the performance deficits expected of finer-grained coverage (e.g., from spending more time covering more loops).
Furthermore, as column 3 in \autoref{tab:perf:piecewise} shows, \platname's best-case performance is \textbf{nearly indistinguishable} from UnTracer's, with performance statistically improved or identical on all but two benchmarks.

\par{\textbf{Versus binary-only always-on tracing}:}
As \autoref{fig:perf} shows, \platname averages \textbf{11.4$\times$}, \textbf{24.1$\times$}, and \textbf{3.6$\times$} the throughput of always-on binary-only tracers QEMU, Dyninst, and RetroWrite, respectively.
Furthermore, we observe that \textbf{all~23} comparisons to \platname yield a statistically significant improvement in \platname's speed over these competing binary-only tracers.

\par{\textbf{Versus source-level always-on tracing}:}
\platname averages \textbf{2.8$\times$} the throughput of AFL's main source-level coverage tracer AFL-Clang.
In only one case (\texttt{nasm}) does \platname face lower a throughput of around 19\%; however, the remaining seven open-source benchmarks see \platname attaining a statistically higher throughput.
Thus, we conclude that \platname's coverage-preserving CGT indeed upholds the speed advantages of CGT---outperforming even the ordinarily-fast source-level tracing.

\vspace{0.1cm}
\stevebox{}{
\small
\textbf{Q2:} 
Coverage-preserving CGT trades-off a negligible amount of speed to attain the highest binary-only code and loop coverage---and still outperforms conventional always-on binary- \emph{and} source-level tracing with over 2--24$\times$ the test case throughput.
}


\subsection{Q3: Bug-finding  Evaluation}
\label{sec:eval:bugs}

We evaluate the crash- and bug-finding effectiveness of coverage-preserving CGT across our 12 benchmarks.
To triage raw crashes into bugs, we apply the popular ``fuzzy stack hashing'' methodology, trimming stack traces to their top-6 entries, and hash each with their corresponding fault address and reported error. 
We make use of the binary-only AddressSanitizer implementation QASan~\cite{fioraldi_fuzzing_2020} to extract crash stack traces and errors.

\subsubsection{\textbf{Unique Bugs and Crashes}}
\autoref{tab:bugs} shows the \platname's mean crash- and bug-finding relative to block-coverage-only CGT UnTracer; and always-on fuzzing coverage tracers QEMU, Dyninst, RetroWrite, and AFL-Clang.
\autoref{fig:bugs} shows the mean unique crashes over time for several benchmarks.
We omit \texttt{lzturbo} and \texttt{rar} as no fuzzing run found crashes in them.

\begin{table*}[!t]
\scriptsize   
    \centering
    
    \begin{tabular}[b]{l | c | c | M{1.3cm}  M{1.3cm} | M{1.2cm}  M{1.2cm}  M{1.2cm}  M{1.2cm} }
        \specialrule{.1em}{0em}{0em} 
        \hline
            \mystrut
            \multirow{2}{*}{\textbf{Identifier}} &
            \multirow{2}{*}{\textbf{Category}} &
            \multirow{2}{*}{\textbf{Binary}} & 
            \multicolumn{2}{c|}{{\scriptsize{\emph{Coverage-guided Tracing}}}} &
            \multicolumn{4}{c}{{\scriptsize{Binary- and Source-level \emph{Always-on Tracing}}}} \\
            
            \mystrut
            & & & 
            \scriptsize\textbf{\cellcolor{white}\platname} &
            \scriptsize\textbf{\cellcolor{white}UnTracer} &
            \scriptsize{\cellcolor{white}\textbf{QEMU}} &
            \scriptsize\textbf{\cellcolor{white}Dyninst} &
            \scriptsize{\cellcolor{white}\textbf{RetroWrite}} & 
            \scriptsize{\cellcolor{white}\textbf{Clang}} \\
            
        \hline\hline
        
        \rowcolor{gray!10} CVE-2011-4517\eat{~\cite{CVE-2011-4517}} & heap overflow & \texttt{jasper} & \cellcolor{gray!30}13.1 hrs  & 18.2 hrs  & \xmarktab & \xmarktab & \xmarktab & 8.70 hrs  \\
		
		GitHub issue \#58-1\eat{-1~\cite{Issue-58}} & stack overflow & \texttt{mjs} & \cellcolor{gray!30}13.3 hrs & 19.0 hrs  & \xmarktab & \xmarktab & 15.30 hrs  & \xmarktab \\

		\rowcolor{gray!10}GitHub issue \#58-2\eat{-2~\cite{Issue-58}} & stack overflow & \texttt{mjs} & \cellcolor{gray!30}13.6 hrs  & 16.4 hrs  & \xmarktab & 22.6 hrs  & \xmarktab & 15.70 hrs  \\

		GitHub issue \#58-3\eat{-3~\cite{Issue-58}} & stack overflow & \texttt{mjs} & \cellcolor{gray!30}5.88 hrs  & 6.80 hrs  & \xmarktab & 14.7 hrs  & \xmarktab & \xmarktab \\

	    \rowcolor{gray!10}GitHub issue \#58-4\eat{-4~\cite{Issue-58}} & stack overflow & \texttt{mjs} & \cellcolor{gray!30}8.60 hrs  & 10.7 hrs  & \xmarktab & 20.1 hrs  & 19.6 hrs & \xmarktab \\

		GitHub issue \#136\eat{~\cite{Issue-136}} & stack overflow & \texttt{mjs} & \cellcolor{gray!30}1.30 hrs  & 7.50 hrs  & \xmarktab & 1.30 hrs  & \xmarktab & \xmarktab \\
		
	    \rowcolor{gray!10}Bugzilla \#3392519\eat{~\cite{3392519}} & \scriptsize{null pointer deref} & \texttt{nasm} & \cellcolor{gray!30}12.1 hrs  & 13.5 hrs  & \xmarktab & \xmarktab & \xmarktab & \xmarktab \\
		
		CVE-2018-8881\eat{~\cite{CVE-2018-8881}} & heap overflow & \texttt{nasm} & \cellcolor{gray!30}5.06 hrs  & 14.6 hrs  & \xmarktab & \xmarktab & \xmarktab & 13.9 hrs  \\

		\rowcolor{gray!10}CVE-2017-17814\eat{~\cite{CVE-2017-17814}} & use-after-free & \texttt{nasm} & \cellcolor{gray!30}3.54 hrs  & 6.31 hrs  & \xmarktab & \xmarktab & \xmarktab & 5.91 hrs  \\

		CVE-2017-10686\eat{~\cite{CVE-2017-10686}} & use-after-free & \texttt{nasm} & \cellcolor{gray!30}3.84 hrs  & 5.40 hrs  & \xmarktab & \xmarktab & \xmarktab & 4.70 hrs  \\

		\rowcolor{gray!10}Bugzilla \#3392423\eat{~\cite{3392423}} & illegal address & \texttt{nasm} & \cellcolor{gray!30}8.17 hrs  & 14.2 hrs  & \xmarktab & \xmarktab & \xmarktab & \xmarktab \\

	    CVE-2008-5824\eat{~\cite{CVE-2008-5824}} & heap overflow & \texttt{sfconvert} & \cellcolor{gray!30}13.1 hrs  & 14.8 hrs  & \xmarktab & 14.3 hrs  & 15.4 hrs  & \xmarktab \\

		\rowcolor{gray!10}CVE-2017-13002\eat{~\cite{CVE-2017-13002}} & stack over-read & \texttt{tcpdump} & \cellcolor{gray!30}8.34 hrs  & 12.5 hrs  & \xmarktab & 13.5 hrs  & 11.5 hrs  & 8.04 hrs  \\

		CVE-2017-5923\eat{~\cite{CVE-2017-5923}} & heap over-read & \texttt{yara} & \cellcolor{gray!30}3.24 hrs  & 5.67 hrs  & 1.87 hrs  & \xmarktab & 9.33 hrs  & 6.19 hrs  \\

		\rowcolor{gray!10}CVE-2020-29384\eat{~\cite{CVE-2020-29384}} & integer overflow & \texttt{pngout} & \cellcolor{gray!30}5.40 min \eat{0.09 hrs}  & 34.5 min \eat{0.58 hrs}  & 18.0 min \eat{0.30 hrs}  & \xmarktab & \xmarktab & \xmarktab \\
		
		CVE-2007-0855\eat{~\cite{CVE-2007-0855}} & stack overflow & \texttt{unrar} & \cellcolor{gray!30}10.7 hrs  & 17.6 hrs  & \xmarktab & \xmarktab & \xmarktab & \xmarktab \\

        \hline
        \multicolumn{4}{l}{\textbf{\platname's Mean Relative Speedup}} & \textbf{52.4\%} & \textbf{48.9\%} & \textbf{41.2\%} & \textbf{43.5\%} & \textbf{32.3\%} \\
    \end{tabular}
    \vspace{-0.0cm}
    \caption{\textbf{\platname's mean bug time-to-exposure} relative to block-coverage-only CGT UnTracer; and conventional always-on coverage tracers QEMU, Dyninst, RetroWrite, and AFL-Clang. \textbf{\xmarktab} = the competing tracer is incompatible with the benchmark or does not uncover the bug.}
    \label{tab:bugtimes}
\vspace{-0.4cm}
\hrulefill{}
\vspace{0.2cm}
\end{table*}

\par\textbf{Versus UnTracer:}
As \autoref{tab:bugs} shows, \platname exposes a mean \textbf{12\%} more bugs than UnTracer.
In conjunction with the plots shown in \autoref{fig:bugs}, we see that coverage-preserving CGT's small sacrifice in speed is completely offset by the much higher number of bugs and crashes found---attaining effectiveness statistically better than or identical to UnTracer on all 12 benchmarks.

\par\textbf{Versus binary-only always-on tracing:}
As expected, \platname's coverage-preserving CGT attains a mean improvement of \textbf{521\%}, \textbf{1193\%}, and \textbf{56\%} in fuzzing bug-finding over always-on binary-only tracers QEMU, Dyninst, and RetroWrite (respectively).
Just as in our performance experiments (\autoref{sec:eval:perf}), \textbf{all 21} comparisons yield a statistically significant improvement for \platname.

\par\textbf{Versus source-level always-on tracing:}
Across all eight open-source benchmarks, \platname achieves a \textbf{46\%} higher bug-finding effectiveness than source-level tracer AFL-Clang, with statistically improved and statistically identical bug-finding on 6/8 and 2/8 binaries (respectively). 
Overall, beating even source-level tracers highlights \platname's value at \emph{binary-only} coverage.

\subsubsection{\textbf{Bug Diversity}}
\label{sec:eval:bugs:diversity}
Following additional triage to map discovered crashes to previously-reported vulnerabilities and bugs, we conduct several case studies to further examine \platname's practicality in real-world bug-finding versus existing tracers.

To determine whether coverage-preserving CGT effectively reveals \emph{many} bugs, or is merely constrained to the same few time after time, we compare the total bugs found by \platname to the best-performing always-on coverage-tracers, RetroWrite (binary-only) and AFL-Clang (source-level).
As \autoref{fig:bugs:venn} shows, despite some overlap, \platname reveals \textbf{1.4$\times$} the unique bugs as RetroWrite and AFL-Clang---with a higher number of bugs that \emph{only} \platname successfully reveals---confirming that coverage-preserving CGT is practical for real-world bug-finding.

\definecolor{vd_z}{HTML}{1B75AE}
\definecolor{vd_u}{HTML}{239C75}
\definecolor{vd_q}{HTML}{DA8D32}
\definecolor{vd_d}{HTML}{1B75AE}
\definecolor{vd_c}{HTML}{D88FC1}

\vspace{-0.1cm}
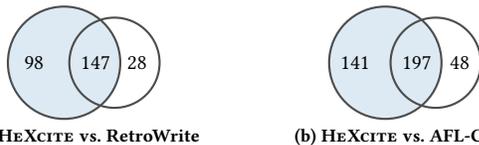
\begin{figure}[!h]
    \small
    \begin{subfigure}[b]{0.24\textwidth}
        \centering
        \begin{tikzpicture}
            \colorlet{circle edge}{black!70}
            \colorlet{circle area}{vd_z!15}
            \tikzset{
                 filled/.style={fill=circle area, thick,inner sep=0pt}, 
                 outline/.style={draw=circle edge, thick,inner sep=0pt}}
            \node (secondcircle) [circle,text width=0.8cm] {};
            \node (firstcircle) [circle,filled,left=-0.6cm of secondcircle,outline,text width=1.5cm] {};
            \draw [outline] (secondcircle) circle (0.6cm);
            \node at ([xshift=-0.4cm]firstcircle) {{98}};
            \node at ([xshift=0.3cm]secondcircle) {28};
            \node at ($(firstcircle)!0.625!(secondcircle)$) {147};
        \end{tikzpicture}
        \vspace{-0.125cm}
        \caption{\textbf{\platname vs. RetroWrite}}
        \label{fig:venn:A}
    \end{subfigure}%
    \begin{subfigure}[b]{0.24\textwidth}
        \centering
        \begin{tikzpicture}
            \colorlet{circle edge}{black!70}
            \colorlet{circle area}{vd_z!15}
            \tikzset{
                 filled/.style={fill=circle area, thick,inner sep=0pt}, 
                 outline/.style={draw=circle edge, thick,inner sep=0pt}}
            \node (secondcircle) [circle,text width=0.8cm] {};
            \node (firstcircle) [circle,filled,left=-0.6cm of secondcircle,outline,text width=1.5cm] {};
            \draw [outline] (secondcircle) circle (0.6cm);
            \node at ([xshift=-0.4cm]firstcircle) {{141}};
            \node at ([xshift=0.32cm]secondcircle) {48};
            \node at ($(firstcircle)!0.625!(secondcircle)$) {197};
        \end{tikzpicture}
        \vspace{-0.125cm}
        \caption{\textbf{\platname vs. AFL-Clang}}
        \label{fig:venn:B}
    \end{subfigure}
    \vspace{-0.7cm}
    
    \caption{\textbf{\platname's total unique bugs found} versus the fastest conventional always-on tracers RetroWrite (binary-only) and AFL-Clang (source-level).}
    \vspace{-0.23cm}
    \hrulefill
    \label{fig:bugs:venn}    
\end{figure}

\subsubsection{\textbf{Bug Time-to-Exposure}}
\label{sec:eval:bugs:times}
We further compare \platname's mean time-to-exposure for 16 previously-reported bugs versus block-only CGT UnTracer; and always-on coverage tracers QEMU, Dyninst, RetroWrite, and AFL-Clang.
As \autoref{tab:bugtimes} shows, \platname accelerates bug discovery by \textbf{52.4\%}, \textbf{48.9\%}, \textbf{41.2\%}, \textbf{43.5\%}, and \textbf{32.3\%} over UnTracer, QEMU, Dyninst, RetroWrite, and AFL-Clang (respectively).
While \platname is not the fastest on every bug, its overall improvement over competing tracers further substantiates the improved fuzzing effectiveness of coverage-preserving CGT.

\vspace{0.1cm}
\stevebox{}{
\small
\textbf{Q3:} 
Coverage-preserving CGT's balance of speed and coverage improves fuzzing effectiveness, revealing more bugs than alternative tracing approaches---in less time.
}


\section{Discussion}
Below we discuss several limitations of coverage-preserving CGT and our prototype implementation, \platname.
			
\subsection{Indirect Critical Edges}
\label{sec:disc:indy}
While resolving \emph{direct} critical edges is straightforward through jump mistargeting or edge splitting (\autoref{sec:covg:edgecovg}), \emph{indirect} critical edges (i.e., indirect jumps/calls/returns) remain a universal problem even for source-level solutions like LLVM's SanitizerCoverage~\cite{the_clang_team_sanitizercoverage_2019}.
Below we discuss several emerging and/or promising techniques for resolving indirect critical edges, and their trade-offs with respect to supporting a binary-level coverage-preserving CGT.

\par\textbf{Block Header Splitting:}
LLVM's SanitizerCoverage supports resolving indirect critical edges whose end blocks have one or more incoming \emph{direct} edges. 
For example, given a CFG with indirect critical edge $\vv{\texttt{ib}}$ (with \texttt{i} having outgoing indirect edges to some other blocks \texttt{x} and \texttt{y}) and direct edge $\vv{\texttt{ab}}$, SanitizerCoverage first cuts block \texttt{b}'s header from its body into two copies, $\texttt{b}_{0i}$ and $\texttt{b}_{0a}$. 
Second, as the indirect transfer's destination is resolved dynamically and thus cannot be statically moved, \texttt{b$_{0i}$}'s location must be pinned to that of the original block \texttt{b}.
Finally, the twin header blocks ($\texttt{b}_{0i}$ and $\texttt{b}_{0a}$) are appended with a direct jump to \texttt{b}'s body, $\texttt{b}_{1}$---effectively splitting the original indirect critical edge $\vv{\texttt{ib}}$ with edges $\vv{\texttt{ib$_{0i}$}}$ and $\vv{\texttt{b$_{0i}$b$_1$}}$; and direct edge $\vv{\texttt{ab}}$ with $\vv{\texttt{ab$_{0a}$}}$ and $\vv{\texttt{b$_{0a}$b$_1$}}$. 
However, the inability to statically alter indirect transfer destinations makes this approach only applicable for indirect critical edges that are the \emph{sole} indirect edge to their end block; i.e., should there be multiple indirect critical edges ($\vv{\texttt{i$_1$b}}$ and $\vv{\texttt{i$_2$b}}$), \emph{at most one} can be split. 

\par\textbf{Indirect Branch Promotion:}
Originally designed as a mitigation for branch target prediction attacks, indirect branch promotion aims to ``rewrite'' indirect transfers as direct:
at runtime, each dynamically-resolved indirect branch target is compared to several statically-encoded candidates, with a conditional jump to each should the comparison match (e.g., \texttt{if(\%eax == foo): jump foo}).
While promotion is applicable to nearly all indirect branches (and hence indirect critical edges), branch target prediction accuracy is \emph{never} guaranteed.
Existing approaches attempt to maximize precision by profiling indirect branches in advance for their ``most probable'' targets, however, fuzzing may expose (and prioritize) new targets previously considered unlikely by profiling.

\par\textbf{Hybrid Instrumentation:}
A third possibility for indirect critical edges is to default back to AFL-style hashing-based edge coverage (\autoref{sec:back:covg}).
While it is impossible to identify each indirect edge's targets accurately, a conservative approach is to instead instrument the set of \emph{all} potential indirect branch targets, as their heuristics are generally well-known (e.g., function entrypoints for indirect calls, and post-call blocks for returns).
We can thus imagine future \emph{target-tailored} CGT approaches balancing fast speed for common-case critical edges with more precise handling (e.g., header splitting, promotion, and hybrid instrumentation) of infrequent ones. 

\subsection{Trade-offs of Hit Count Coverage}
Hit counts measure fuzzing exploration progress in loops and cycles, but as with any coverage metric, their implementation must carefully balance precision and speed to support effective bug-finding.
Two considerations central to hit count coverage implementations are (1) the size and number of bucket ranges; and (2) the frequency at which hit counts are tracked.
We discuss both of these below.

\par\textbf{Bucket Granularity:}
Our current implementation of bucketed unrolling (\autoref{sec:impl:bunroll}) mimics the hit count tracking of conventional fuzzers by injecting conditional checks against eight bucket ranges (0--1, 2, 3, 4--7, 8--15, 16--31, 32--127, 128+).
However, \emph{these eight} bucket ranges are merely an artifact of AFL's original implementation (each hashed edge is mapped to an 8-bit index in its coverage bitmap).
Adding \emph{more} buckets makes it possible to track more subtle changes in loop iteration counts, while using \emph{fewer} buckets trades-off this level of introspection for higher fuzzing throughput.
While it is unclear which bucket ranges achieve the \emph{best} balance of speed and coverage with respect to bug-finding, we expect that future research will address these unanswered questions and more.

\par\textbf{Frequency of Tracking:}
How often hit counts are tracked further influences fuzzing exploration and bug-finding.
Conventional \emph{exhaustive} (per-edge) hit counts shed light on frequencies of cycle subpaths (e.g., how many times a loop \texttt{break} is taken), but risk saturating a fuzzer's search space with redundant or noisy paths. 
Bucketed unrolling instead trades-off coverage exhaustiveness for speed by restricting hit count tracking to only a \emph{subset} of the program state (e.g., loop iteration counters).
While our analysis of the bugs exclusively found by exhaustive hit counts (\autoref{fig:venn:B}) reveals that none are outside the reach of \platname, we expect that future work will explore adapting \emph{selective} and \emph{synergistic} hit count schemes to better cover complex loops, cycles, and compiler optimizations at high speed.

\subsection{Improving Performance}
The fuzzing-oriented binary transformation platform currently utilized in \platname, ZAFL~\cite{nagy_breaking_2021}, adopts a code layout algorithm that rewrites all direct jumps to have 32-bit PC-relative signed displacements.
While this is well-suited to our implementation of zero-address jump mistargeting (\autoref{sec:impl:jmpmis})---enabling virtually every conditional jump in the program's address space to be mistargeted to \texttt{0x00}---32-bit displacements accumulate more runtime overhead over 8--16-bit displacements.
As ZAFL has experimental code layouts that instead prioritize smaller displacements, we thus envision potential for faster ``hybrid'' mistargeting schemes that coalesce both zero-address \emph{and} embedded interrupt styles.

\subsection{Supporting Other Software \& Platforms}
Our current coverage-preserving CGT prototype, \platname, supports 64-bit Linux C and C++ binaries.
Extending support to other software characteristics (e.g., 32-bit) or platforms (e.g., Windows) requires retooling of its underlying static binary rewriting engine.
However, as this component is orthogonal to the fundamental principles of coverage-preserving CGT, we expect that \platname will capitalize on future engineering improvements in static rewriting to bring accelerated fuzzing to the broader software ecosystem. 

\section{Related Work}
We discuss recent efforts to improve binary-only fuzzing performance that are orthogonal to coverage-preserving CGT: (1) faster instrumentation, (2) less instrumentation, and (3) faster execution.

\subsection{Faster Instrumentation}
As binary fuzzing effectiveness depends heavily on maintaining fast coverage tracking, a growing body of research is targeting instrumentation-side optimizations.
Efforts to improve dynamic translation-based instrumentation (e.g., AFL-QEMU~\cite{zalewski_american_2017}, DrAFL~\cite{shudrak_drafl_2019-1}, UnicornAFL~\cite{voss_afl-unicorn_2019}) generally focus on simplifying or expanding the caching of translated code~\cite{biondo_improving_2018}; 
while those using static rewriting (e.g., ZAFL~\cite{nagy_breaking_2021}, Dyninst~\cite{paradyn_tools_project_dyninst_2018}, RetroWrite~\cite{dinesh_retrowrite_2020}) tackle various challenges related to generated code performance.
Though our coverage-preserving CGT prototype, \platname, currently leverages the ZAFL rewriter, we believe that future advances in binary instrumentation will enable it to achieve performance even closer to native speed.

\subsection{Less Instrumentation}
Another way to reduce the footprint of coverage tracking is to eliminate needless instrumentation from the program under test.
While most other control-flow-centric approaches only exist in compiler instrumentation-based implementation (e.g., dominator trees~\cite{agrawal_dominators_1994}, INSTRIM~\cite{hsu_instrim_2018}, CollAFL~\cite{gan_collafl_2018}), their principles are well-suited to binary-only fuzzing.
A recent fork of AFL-Dyninst~\cite{heuse_afl-dyninst_2018} omits instrumentation from blocks preceded by unconditional direct transfer, as their coverage is directly implied by their ancestor's.
In addition to accelerating execution of \platname's tracer binary, we see the potential for such control-flow-centric analyses to help determine how \platname's control-flow-altering transformations (e.g., bucketed unrolling) should optimally be applied.

\subsection{Faster Execution}
Besides instrumentation, execution is itself a bottleneck to fuzzing, as faster execution enables more test cases to be run on the target program in less time.
Most modern binary-only fuzzing efforts have abandoned slow process creation-based execution for faster snapshotting, leveraging cheap copy-on-write cloning to rapidly initiate target execution from a pre-initialized state (e.g., AFL's forkserver~\cite{zalewski_american_2017}). 
Xu et al.~\cite{xu_designing_2017} achieve even faster snapshotting through fuzzing-optimized Linux kernel extensions.
The recent technique of persistent/in-memory execution offers higher speed by restricting execution to only a pre-specified target program code region (essentially interposing a loop), and is gaining support among popular binary-only fuzzing toolchains (e.g., WinAFL~\cite{google_project_zero_winafl_2016}, AFL-QEMU, UnicornAFL).
Many efforts are also exploring the benefits of amortizing fuzzing execution speed through parallelization; off-the-shelf binary-only fuzzers like AFL~\cite{zalewski_american_2017} and honggFuzz~\cite{swiecki_honggfuzz_2018} support parallelization out-of-the-box, and recent work by Falk~\cite{falk_vectorized_2018} achieves even faster speed by leveraging vectorized instruction sets. 
As execution and coverage tracking work hand-in-hand during fuzzing, we view such accelerated execution mechanisms as complementary to \platname's accelerated coverage tracking. 

\section{Conclusion}
Coverage-preserving Coverage-guided Tracing extends the principles behind CGT's performance-maximizing, waste-eliminating tracing strategy to the finer-gained coverage metrics it is not naturally supportive of: edge coverage and hit counts.
We introduce program transformations that enhance CGT's introspection capabilities while upholding its minimally-invasive nature; and show how these techniques improve binary-only fuzzing effectiveness over conventional CGT, while keeping an orders-of-magnitude performance advantage over the leading binary-only coverage tracers.

Our results reveal it is finally possible for today's state-of-the-art coverage-guided fuzzers to embrace the acceleration of CGT---without sacrificing coverage.
We envision a new era in software fuzzing, where synergistic and target-tailored approaches will maximize \emph{common-case} performance with \emph{infrequent-case} precision.

\section*{Acknowledgement}
We thank our shepherd Jun Xu and our reviewers for helping us improve the paper.
We also thank Peter Goodman and Trail of Bits for assisting us with binary-to-LLVM lifting.
This material is based upon work supported by the Defense Advanced Research Projects Agency under Contract No. W911NF-18-C-0019 
and the National Science Foundation under Grant Nos. 1650540 and 2115130.

\bibliographystyle{plainurl}
\bibliography{references}

\end{document}